%% file: main.tex
\documentclass[twocolumn]{aastex63}
\usepackage{fp}
\usepackage{bm}
\usepackage{xcolor}
\DeclareSymbolFont{matha}{OML}{txmi}{m}{it}
\DeclareMathSymbol{\varv}{\mathord}{matha}{118}

\shorttitle{Photometrically-Classified SLSNe}
\shortauthors{Hsu et al.}

\input{affil.tex}
\defcitealias{Planck_2016}{Planck Collaboration 2016}

\begin{document}

\title{Photometrically-Classified Superluminous Supernovae from the Pan-STARRS1 Medium Deep Survey: A Case Study for Science with Machine Learning-Based Classification}

\correspondingauthor{Brian Hsu}
\email{brianhsu@college.harvard.edu}

\author[0000-0002-9454-1742]{Brian~Hsu}
\CfA

\author[0000-0002-0832-2974]{Griffin~Hosseinzadeh}
\UA

\author[0000-0002-5814-4061]{V.~Ashley~Villar}
\affiliation{Department of Astronomy \& Astrophysics, The Pennsylvania State University, University Park, PA 16802, USA}
\affiliation{Institute for Computational \& Data Sciences, The Pennsylvania State University, University Park, PA 16802, USA}
\affiliation{Institute for Gravitation and the Cosmos, The Pennsylvania State University, University Park, PA 16802, USA}

\author[0000-0002-9392-9681]{Edo~Berger}
\CfA
\IAIFI

\begin{abstract}
With the upcoming Vera C.~Rubin Observatory Legacy Survey of Space and Time (LSST), it is expected that only $\sim 0.1\%$ of all transients will be classified spectroscopically. To conduct studies of rare transients, such as Type I superluminous supernovae (SLSNe), we must instead rely on photometric classification.  In this vein, here we carry out a pilot study of SLSNe from the Pan-STARRS1 Medium-Deep Survey (PS1-MDS) classified photometrically with our {\tt SuperRAENN} and {\tt Superphot} algorithms.  We first construct a sub-sample of the photometric sample using a list of simple selection metrics designed to minimize contamination and ensure sufficient data quality for modeling.  We then fit the multi-band light curves with a magnetar spin-down model using the Modular Open-Source Fitter for Transients ({\tt MOSFiT}). Comparing the magnetar engine and ejecta parameter distributions of the photometric sample to those of the PS1-MDS spectroscopic sample and a larger literature spectroscopic sample, we find that these samples are overall consistent, but that the photometric sample extends to slower spins and lower ejecta masses, which correspond to lower luminosity events, as expected for photometric selection.  While our PS1-MDS photometric sample is still smaller than the overall SLSN spectroscopic sample, our methodology paves the way to an orders-of-magnitude increase in the SLSN sample in the LSST era through photometric selection and study.
\end{abstract}

\section{Introduction} 
\label{sec:intro}

Hydrogen-poor (Type I) superluminous supernovae (hereafter SLSNe) are a rare sub-class of core-collapse supernovae (CCSNe) that radiate $\sim 10-100$ times more energy in the UV/optical than typical CCSNe, and generally exhibit longer durations and hotter continuum spectra \citep[e.g.,][]{Chomiuk_2011,Quimby_2011,Nicholl_2015b,Inserra_2017,Lunnan_2018,DeCia_2018}.  SLSNe account for only $\sim 0.1\%$ of the volumetric CCSN rate \citep{Quimby_2018,Frohmaier_2021}, but in magnitude-limited optical surveys they account for $\sim 2\%$ of all transients \citep{Perley_2020,Gomez_2021} thanks to their high luminosity. SLSNe are classified spectroscopically based on the lack of  hydrogen Balmer lines, the presence of a blue continuum, and unique early time ``W''-shaped \ion{O}{2} absorption lines at $\sim 3600-4600$ \AA\ \citep[e.g.,][]{Lunnan_2013,Mazzali_2016,Quimby_2018,Nicholl_2021}. 

Several mechanisms have been proposed to power SLSNe, but a magnetar central engine model \citep{Kasen_Bildsten_2010,Woosley_2010,Dessart_2012,Metzger_2015,Nicholl_2017b} has had the most success in explaining both the light curves and spectra of the SLSN population. This model accounts for the broad range of peak luminosities and timescales (e.g., \citealt{Nicholl_2017b,Blanchard_2020}), for the early UV/optical spectra (e.g., \citealt{Nicholl_2017c}), for the nebular phase spectra (e.g., \citealt{Nicholl_2016b,Nicholl_2019,Jerkstrand_2017}), and for the power law decline rates observed in SN\,2015bn and SN\,2016inl at $\gtrsim 10^3$ d \citep{Nicholl_2018,Blanchard_2021}. Additional support for a magnetar engine comes from the low metallicity host galaxies of SLSNe, which most closely resemble the hosts of long-duration gamma-ray bursts, another rare population of CCSNe that are likely powered by a central engine \citep{Lunnan_2014,Perley_2016}. While the magnetar engine model can explain the plethora of SLSN properties, other mechanisms have also been proposed to explain some SLSN properties; for example, \citet{Chen_2022b} recently argued that the light curves of at least some SLSNe from the Zwicky Transient Facility (ZTF; \citealt{Bellm_2019}) can be explained equally well with a combination of circumstellar interaction (CSM) and Ni$^{56}$ decay. Furthermore, \cite{Hosseinzadeh_2021} also explored ejecta-CSM interaction as a potential source for post-peak undulations in SLSN light curves.

With ongoing and upcoming wide-field optical surveys, including in particular the Vera C.~Rubin Observatory Legacy Survey of Space and Time \citep[LSST;][]{Ivezic_2019}, only a small fraction of SNe are being classified spectroscopically ($\sim 10\%$ currently, and $\sim 0.1\%$ anticipated for LSST; \citealt{Villar_2020}).  This impacts the ability to advance the study of rare SN classes, such as SLSNe, in particular. As shown by \citet{Villar_2018}, LSST may yield $\sim 10^4$ SLSNe per year to $z\sim 3$ (of which at least $\sim 20\%$ will have well measured physical properties), but identifying these events requires {\it photometric} classification.

Recently, we presented two machine learning-based SN photometric classification pipelines, {\tt SuperRAENN} \citep{Villar_2020} and {\tt Superphot} \citep{Hosseinzadeh_2020}, trained on 2315 SN-like transients from the Pan-STARRS1 Medium Deep Survey \citep[PS1-MDS;][]{Huber_2017}. Both classifiers use multiple SN classes, including in particular SLSNe. {\tt SuperRAENN} combines a novel unsupervised recurrent autoencoder neural network (RAENN) with a random forest classifier for a semi-supervised algorithm. {\tt Superphot} utilizes a random forest approach based on flexible analytic model fits to the light curves and their resulting parameters. 

Here, as a demonstration of the type of approach and analysis that will be essential in the LSST era, we explore and study for the first time, the {\it photometrically-classified} SLSNe from the Pan-STARRS1 Medium Deep Survey \citep[PS1-MDS,][]{Huber_2017}, as identified by {\tt SuperRAENN} and {\tt Superphot}.  We first explore how to effectively construct a pure and well-measured subset of SLSNe from a photometrically-classified sample (\S\ref{sec:data}).  We then model the light curves of the photometrically-classified SLSNe with the same magnetar engine model previously used to study spectroscopically-classified SLSNe (using {\tt MOSFiT}, \citealt{Guillochon_2018}; \S\ref{sec:model}). Finally, we compare the resulting parameter distributions to those of the spectroscopically-classified PS1-MDS SLSNe, as well as to the overall sample of spectroscopically-classified SLSNe (\S\ref{sec:analysis}). 

Throughout the paper, we assume a flat $\Lambda$CDM cosmology with $\Omega_\mathrm{m}=0.308$ and $H_0=67.8\ \text{km}\ \text{s}^{-1}\ \text{Mpc}$, based on the Planck 2015 results \citepalias{Planck_2016}. We correct all photometry for Milky Way extinction using \citet{Schlafly_Finkbeiner_2011} and follow the extinction law of \citet{Fitzpatrick_1999} with $R_V=3.1$.

\section{Sample Construction} 
\label{sec:data}

\input{cut_seq}

The data used in this paper are from the PS1-MDS. We refer the reader to \citet{Chambers_2016} for details of the PS1 survey telescope and PS1-MDS observing strategy, and to \citet{Villar_2020} and \citet{Hosseinzadeh_2020} for the definition of the overall sample of SN-like transients and their light curves, description of the sub-sample of spectroscopically-classified events, the photometric classification approaches and results, all relevant data (including photometry and host galaxy redshifts), and complete descriptions of the algorithms and training processes. 

In this paper we focus on the sample of photometrically-classified SLSNe.\footnote{Both classification pipelines are open-source and available via GitHub: {\url{https://github.com/villrv/SuperRAENN}} and {\url{https://github.com/griffin-h/superphot}}.} Using \texttt{SuperRAENN} \citep{Villar_2020} and \texttt{Superphot} \citep{Villar_2018,Hosseinzadeh_2020} we photometrically classified 58 and 37 SLSNe, respectively, using the same training set of 557 spectroscopically-classified SNe, which includes 17 SLSNe that were studied in \citet{Lunnan_2018}. Here, we adopt the class with the highest probability as the predicted SN type for each transient. 

Combining all transients classified by the two algorithms as SLSNe, and accounting for 28 classified as SLSNe by both, we obtain an initial sample with 67 photometrically-classified SLSNe. To further evaluate and potentially cull the photometric sample, we investigate several post-classification selection criteria.  We find three effective criteria that help to reduce the sample contamination and lead to events with sufficient data to enable robust modeling. Furthermore, we apply an additional quality cut post-modeling based on model convergence. The criteria and their effects on the sample size are summarized in Table~\ref{tab:seq}, and we discuss them in detail below.

\subsection{Active Galactic Nuclei Host Galaxies}

Prior to applying our algorithms to the sample of PS1-MDS SN-like transients, we systematically excluded light curves with long-term variability to avoid contamination from active galactic nuclei (AGN). Still, some large AGN flares with little other variability over the 4.5 year time-span of the survey could survive this preliminary qualitative cut and eventually be classified as SLSNe.  In particular, \citet{Hosseinzadeh_2020} find that 14 photometrically-classified SLSNe with host galaxy spectra that exhibit broad AGN lines are located within $1''$ of the host center\footnote{These transients are PSc000478, PSc010120, PSc010186, PSc020026, PSc030013, PSc052281, PSc110163, PSc130394, PSc130732, PSc350614, PSc390545, PSc400050, PSc480585, and PSc550061.}. While these could in principle be SLSNe located indistinguishably close to an AGN, they are more likely large AGN flares or tidal disruption events, neither of which is a classification category in \texttt{SuperRAENN} and \texttt{Superphot}. Eliminating these events results in a combined sample of 53 events (Table~\ref{tab:seq}, row 2). 

\subsection{Classification Confidence}

Our initial sample requires that the highest classification probability be assigned as SLSN. However, given the number of classification categories, this does not necessarily mean that the classification confidence is high.  \citet{Hosseinzadeh_2020} and \citet{Villar_2020} show that increasing the classification confidence threshold to $p\gtrsim 0.75$ leads to higher purity\footnote{Purity refers to the fraction of a given photometric class that belongs to the equivalent spectroscopic class \citep{Hosseinzadeh_2020}.} across the full range of classes, at the expense of sample completeness. Here we apply a classification confidence threshold of $p_{\rm SLSN}\ge 0.5$ as a compromise between purity and sample size (which corresponds to a purity of $\approx 0.78$, see \citealt{Hosseinzadeh_2020}). This selection cut reduces the sample size from 53 to 36 events (Table~\ref{tab:seq}, row 3).

\subsection{Number of Light Curve Data Points}

Both the classification confidence and the ability to meaningfully model the light curves with {\tt MOSFiT} (\S\ref{sec:model_desc}) are affected by the number of light curve data points; namely, the number of data points relates to the ability to constrain the {\tt MOSFiT} models and return statistically meaningful posterior distributions. Here we set a threshold of $\ge 11$ data points total across the four observed filters ($griz$) to match the number of model free parameters\footnote{One parameter is set to have a constant value, leaving us with 11 free parameters; see \S\ref{sec:model_desc}.}. This selection cut reduces the sample size from 36 to 24 (Table~\ref{tab:seq}, row 4). 

\subsection{Model Convergence}

The aforementioned selection criteria are applied prior to modeling. After all three criteria are applied, we model the 24 photometrically-classified SLSNe with a magnetar central engine model, implemented in {\tt MOSFiT}. Although we have reduced our sample to identify only events with a sufficient number of data points and high confidence as SLSNe, light curves with marginal detections or potentially misclassified events could in principle survive the above pre-modeling selection metrics. Therefore, we include an additional cut based on the model convergence factor as measured by calculating the Gelman-Rubin statistics, or potential scale reduction factor (PSRF; \citealt{Gelman_1992}), which estimates the extent to which the full parameter space has been explored in our MCMC models. \cite{Brooks_Gelman_1998} suggests that PSRF $<1.2$ provides reliable convergence, but we set a stricter threshold of PSRF $<1.1$ as done in \citet{Nicholl_2017b} and \citet{Hsu_2021}, which is also the termination value for our models (see \S\ref{sec:model_desc}). This post-modeling selection cut reduces our sample size from 24 to 19 (Table~\ref{tab:seq}, row 5). 

Our final photometric sample consists of 10 events classified as SLSNe by both algorithms, with the remaining 9 classified as SLSN by either {\tt SuperRAENN} or {\tt Superphot}. See Table~\ref{tab:classifications} for the predicted SN type of each transient in our final sample and their respective classification confidence.

\input{classifications}

\subsection{Justification of Our Choices}

\begin{figure}
    \centering
    \includegraphics[width=\columnwidth]{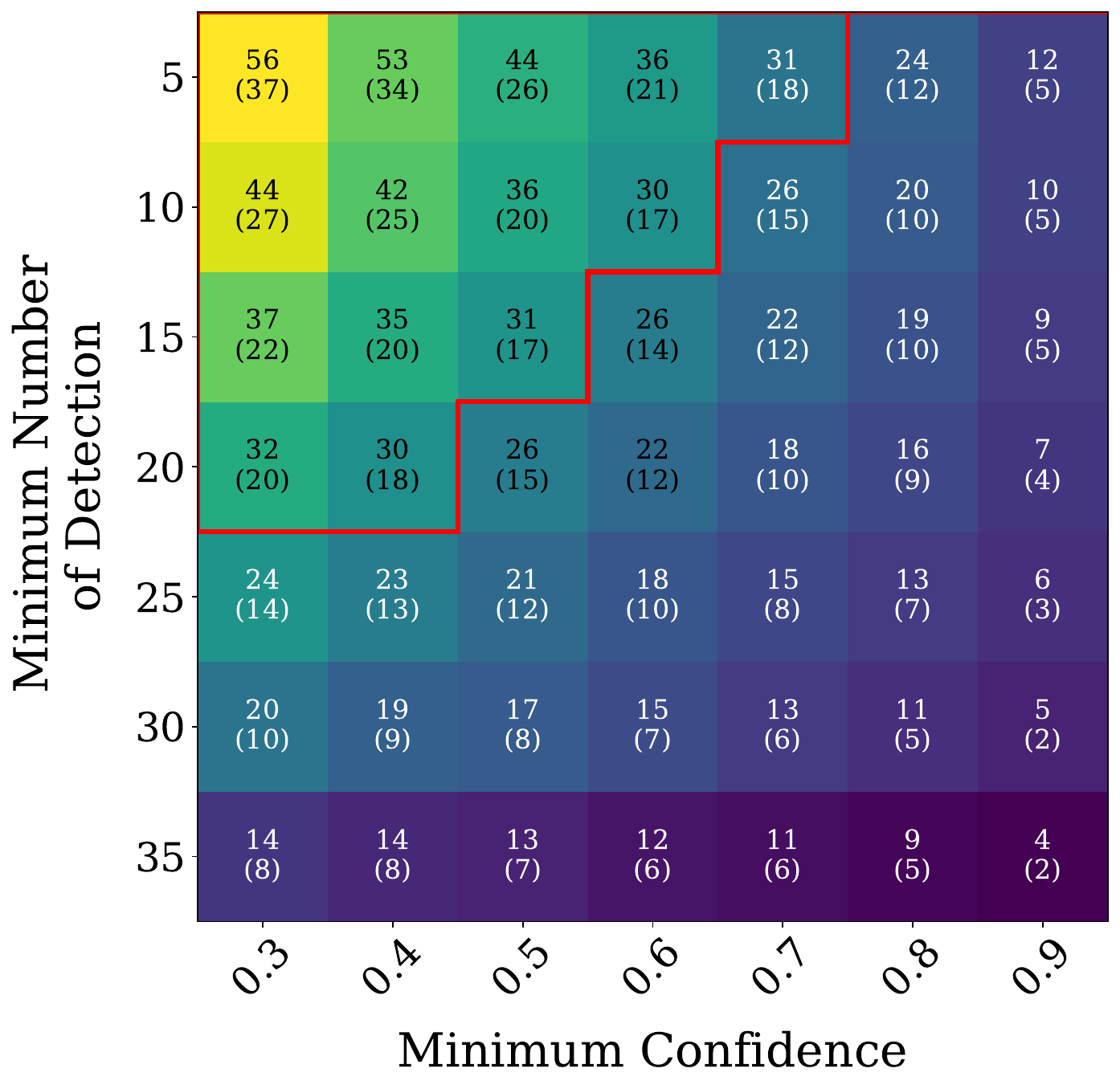}
    \caption{Matrix showing the effect of varying the minimum classification confidence and the minimum number of light curve data points across all 4 filters. The top number in each cell indicates the total number of events (out of 67) that satisfies both thresholds. The numbers in parentheses indicate the final sample size after removing AGN hosts and events with non-converged models. The region outlined in red marks the boundary for combinations that result in a comparable sample size ($\ge17$) to the PS1-MDS spectroscopic sample.}
    \label{fig:cut_mat}
\end{figure}

In Figure~\ref{fig:cut_mat} we show the combined effects on the final sample size of varying the minimum classification confidence and the number of data points; we use this as a guide such that our final sample consists of events with sufficient confidence level and data points to obtain a robust model. In each cell we show the number of events that survive each pair of minimum threshold for confidence and number of detections, and we quote the final sample size after applying both the AGN and convergence cuts in parentheses. To extract a comparable sample size to the PS1-MDS spectroscopic sample (17 events) that will return statistically meaningful results, we outline in Figure~\ref{fig:cut_mat} the combinations of minimum confidence and detection thresholds that produce a minimal final sample size $\ge 17$. We find that our choice of minimum confidence ($\ge 0.5$) and number of detection ($\ge 11$) falls within the outlined region, indicating that our selection criteria are reasonable and justified.

\section{Magnetar Model Fits}
\label{sec:model}

\subsection{Brief Description of the Model}
\label{sec:model_desc}

We fit the optical light curves of the 19 photometrically-classified SLSNe (selected as described in \S\ref{sec:data}) using the Modular Open-Source Fitter for Transients \citep[\texttt{MOSFiT};][]{Guillochon_2018} with the magnetar spin-down model described in \citet{Nicholl_2017b}. \texttt{MOSFiT} is an open-source, \texttt{Python}-based light curve fitting package that employs a Markov chain Monte Carlo (MCMC) algorithm to fit a one-zone, grey-opacity analytical model to multi-band light curves \citep{Nicholl_2017b}.

The magnetar model has 12 free parameters, of which 8 are nuisance parameters that we marginalize over to obtain the 4 key physical parameters related to the ejecta and engine properties. We fix one of the nuisance parameters, the angle $\theta_{PB}$ between the magnetic field and the rotational axis of the magnetar, to $90^{\circ}$ as this ensures that the derived $B$-field strength is a lower limit \citep[following][]{Nicholl_2017b}. The nuisance parameters, $\kappa$, $\kappa_{\gamma}$, $M_{\rm NS}$, $n_{\rm H, host}$, are not well-constrained by the model. Events with sufficient late-time observations may constrain the $\gamma$-ray opacity $\kappa_{\gamma}$, but this is not the case for our sample. The neutron star mass is degenerate with the spin period and magnetic field strength but is not well constrained. The explosion time, $t_{\rm exp}$, is the time between explosion and first observation in the pure magnetar model. The main parameters that constrain the observed properties of a SLSN are the neutron star's initial spin period, $P$, magnetic field strength, $B$, ejecta mass, $M_{\rm ej}$, and ejecta velocity, $\varv_{\rm ej}$ (the latter two can be combined to determine the kinetic energy, $E_K$). The model parameters and their priors are listed in Table~\ref{tab:priors}.

\input{priors}

For each light curve fit, the first 10,000 iterations are used to burn in the ensemble, during which minimization is employed periodically as the ensemble converges to the global optimum; the remainder of the run-time is used to sample the posterior probability distribution. Convergence is measured by calculating the PSRF, and we terminate our fits when PSRF $<1.1$. Most events typically require 30,000--60,000 iterations to reach convergence, depending on the number of data points and the scatter around our model. 

\subsection{Light Curve Fits}
\label{sec:fits}

\begin{figure*}[t!]
    \centering
    \includegraphics[width=\textwidth]{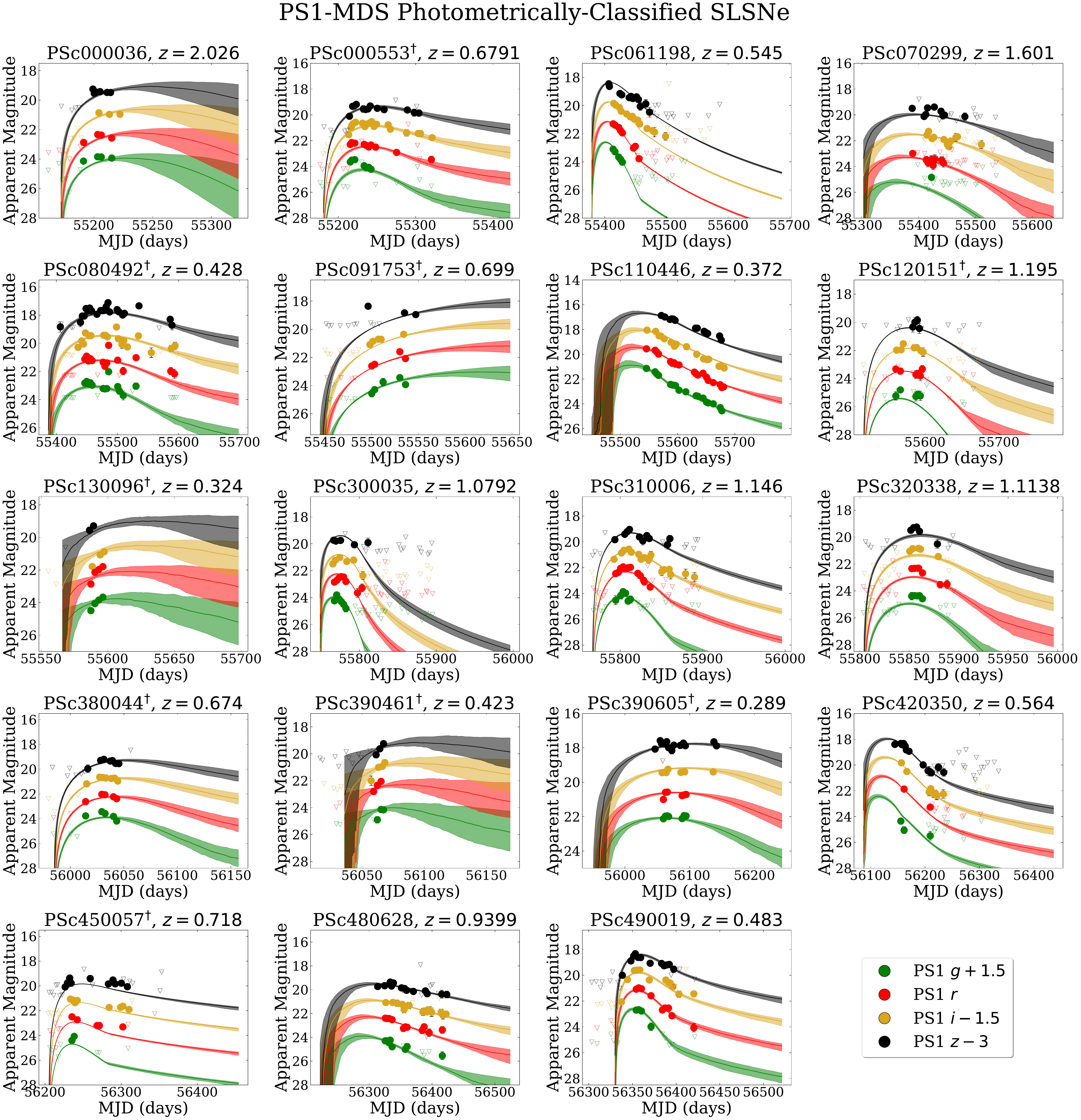}
    \caption{Multiband extinction-corrected apparent magnitude light curves of the 19 PS1-MDS photometrically-classified SLSNe, along with our magnetar model fits using \texttt{MOSFiT}. The name of each transient and its spectroscopic host galaxy redshift are quoted on top of each panel. The different filters are shifted for clarity, as indicated in the legend. Open triangles indicate $3\sigma$ upper limits, while solid circles indicate detections. The solid lines and shaded regions indicate the median model and $1\sigma$ ranges. Events classified as SLSNe by only one classifier are marked with daggers.}
    \label{fig:phot_fits}
\end{figure*}

\begin{figure*}[t!]
    \centering
    \includegraphics[width=\textwidth]{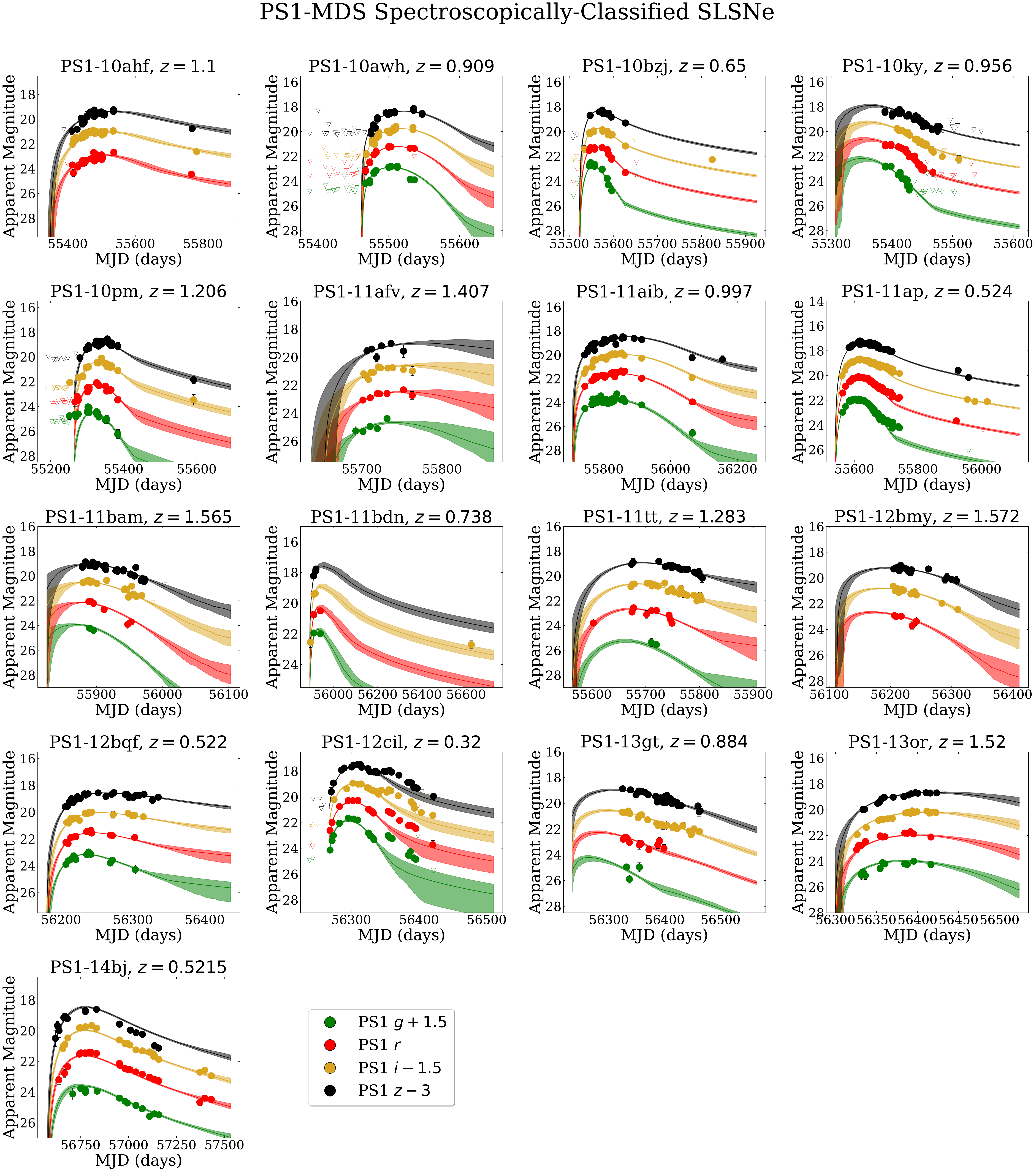}
    \caption{Same as Figure~\ref{fig:phot_fits} but for the PS1-MDS spectroscopically-classified SLSNe. The light curves for PS1-11ap and PS1-12cil are from \cite{Hosseinzadeh_2021}, modeled without the post-peak pumps.}
    \label{fig:spec_fits}
\end{figure*}

In Figure~\ref{fig:phot_fits}, we show the magnetar model light curve fits for the 19 PS1-MDS photometrically-classified SLSNe. The shaded regions are the {\tt MOSFiT} light curve fits, where the upper and lower bounds are 1$\sigma$ uncertainties calculated from the 120 MCMC walkers, while the solid light curves are based on the parameter medians. To allow for a proper comparison with the PS1-MDS spectroscopic sample, we also show in Figure~\ref{fig:spec_fits} the 17 PS1-MDS spectroscopically-classified SLSNe (which were used in the classification training samples). Previous studies \citep{Nicholl_2017b,Villar_2018,Blanchard_2020} have already modeled all but one of these events (PS1-12cil) in the same manner as this work. Two peculiar events from the PS1-MDS spectroscopically-classified sample, PS1-11ap and PS1-12cil, exhibit post-peak undulations \citep[e.g.,][]{Inserra_2013,Nicholl_2014,Inserra_2017,Nicholl_2016a,Hosseinzadeh_2021}. Since our {\tt MOSFiT} model does not account for these ``bumps'', we replace the model of PS1-11ap from \citet{Blanchard_2020} with the version presented in \citet{Hosseinzadeh_2021}, which converts these bumps into upper limits prior to fitting. We also include the model for PS1-12cil from \citet{Hosseinzadeh_2021} to complete the PS1-MDS spectroscopic SLSN sample. For illustrative purposes, we extrapolate all light curves (both photometric and spectroscopic samples) back to the inferred explosion time and forward 100 days after the last detection.

\input{phys_params}

Overall, we find that the model fits the observed light curves well, and is better constrained for events with more extensive data. The resulting median values and $1\sigma$ uncertainties for the four main physical parameters ($P$, $B$, $M_{\rm ej}$, $\varv_{\rm ej}$), calculated based on the posterior probability distributions from 120 MCMC walkers are summarized in Table~\ref{tab:phys_params}. Our model includes an intrinsic scatter term, $\sigma$, that attempts to model white, systematic scatter not captured by our statistical uncertainties.

The sample median values and associated $1\sigma$ ranges of the four key model parameters, along with the kinetic energy\footnote{Our model assumes the analytic density profile described in \citet{Margalit_2018}. For a homogeneous density profile, the kinetic energy is given by $E_K=\frac{3}{10}M_{\rm ej}{\varv_{\rm ej}}^2$.}, $E_K=\frac{1}{2}M_{\rm ej}{\varv_{\rm ej}}^2$, for the PS1-MDS photometric and spectroscopic samples are listed in Table~\ref{tab:median}. We also list in Table~\ref{tab:median} the values for a larger SLSN compilation sample, which includes 82 spectroscopically-classified SLSNe (81 from \citealt{Hsu_2021} plus PS1-12cil from \citealt{Hosseinzadeh_2021}), as a comparison sample. Comparing these samples, we find that the PS1-MDS photometric sample displays somewhat slower spins, higher $B$-field values, and lower ejecta masses as compared to the other two samples. However, the values are in good agreement within the $1\sigma$ ranges. 

\input{median}

\section{Sample Properties}
\label{sec:analysis}

\subsection{Observational Properties}
\label{sec:obs_prop}

\begin{figure}
    \centering
    \includegraphics[width=\columnwidth]{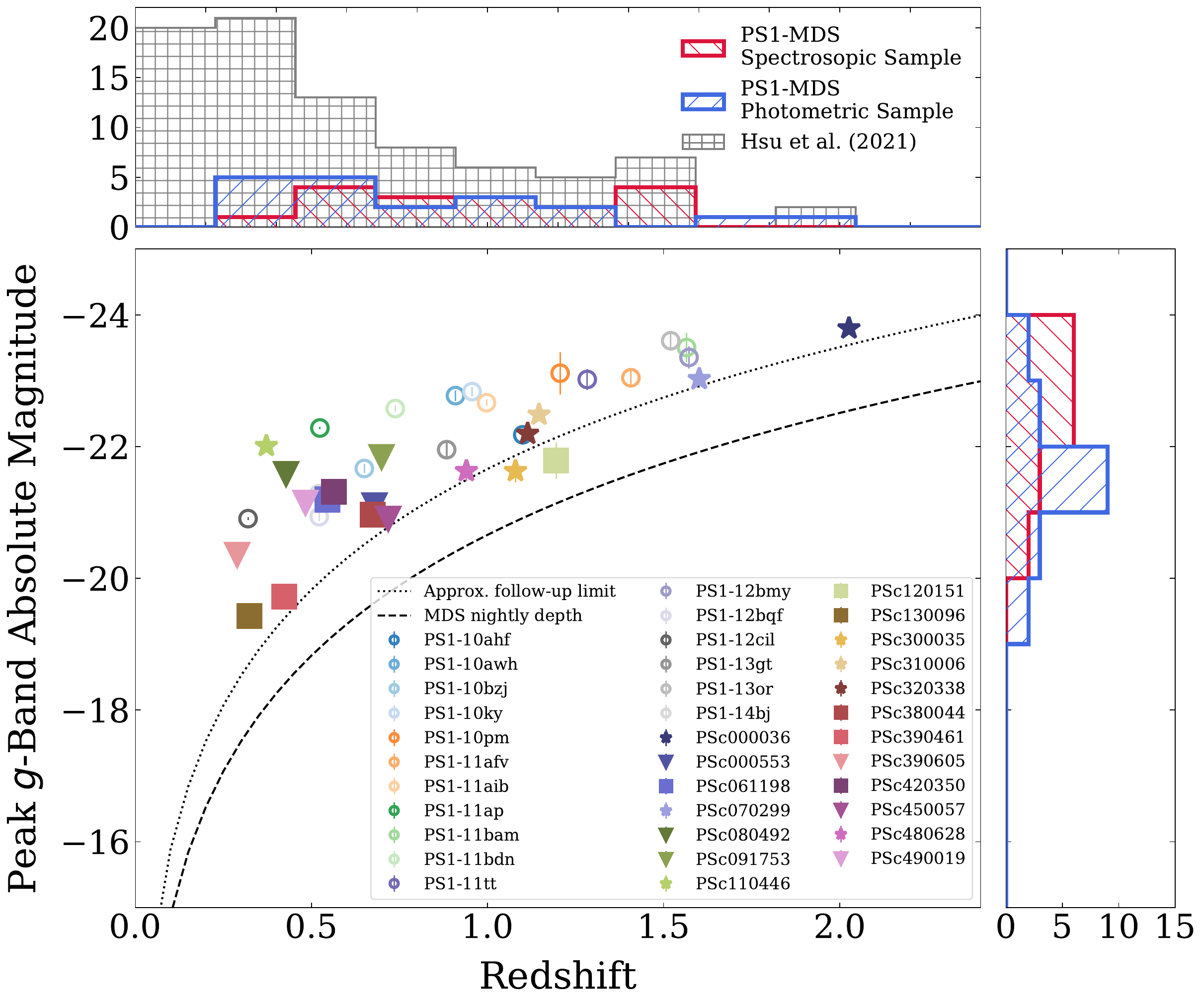}
    \caption{Peak absolute rest-frame $g$-band magnitude versus redshift for the PS1-MDS SLSNe (solid squares: photometrically classified by {\tt Superphot} only; solid triangles: photometrically classified by {\tt SuperRAENN} only; solid stars: photometrically classified by both {\tt SuperRAENN} and {\tt Superphot}; open circles: spectroscopically classified). The top panel shows the redshift distribution of the PS1-MDS photometric (blue) and spectroscopic (red) samples compared to the SLSN compilation from \citet{Hsu_2021} (82 spectroscopic SLSNe, including the 17 from PS1-MDS), while the right panel shows the distributions of peak magnitudes for the PS1-MDS samples. Each peak magnitude is corrected for Galactic extinction and includes a cosmological $K$-correction factor of $2.5\log(1+z)$.}
    \label{fig:mag_z}
\end{figure}

The PS1-MDS samples (both spectroscopic and photometric) collectively span a wide range of redshifts, $z\approx 0.3-2$. To properly compare the observational properties of the PS1-MDS SLSNe, we correct their observed peak apparent magnitudes to a single rest-frame filter. Since we do not have a complete set of spectra for the spectroscopic sample, and by definition no spectra for the photometric sample, we do not apply a complete $K$-correction; instead, we apply only a cosmological $K$-correction factor of $2.5\log_{10}(1+z)$ to the peak magnitude in the band closest to the rest-frame $g$-band for each event and correct for Milky Way extinction. We plot the resulting peak $g$-band absolute magnitudes as a function of redshift in Figure~\ref{fig:mag_z}. The PS1-MDS spectroscopic sample spans a range of $\approx -20.5$ to $\approx -22.6$, while the photometric sample spans a wider range of $\approx -18.7$ to $\approx -22.6$. As expected, lower luminosity SLSNe are restricted to lower redshift ($z\lesssim 0.5$), while higher luminosity events are distributed to higher redshift ($z\approx 2$). The spectroscopic sample is intrinsically more luminous, with a median peak magnitude of $-22$ as compared to $-20.8$ for the photometric sample. 

We also plot in Figure~\ref{fig:mag_z} the per-visit PS1-MDS limiting magnitude of $\approx 23.3$ \citep{Villar_2020}, as well as the effective spectroscopic follow-up limit of $\approx 22.5$ \citep{Lunnan_2018}. The majority of the photometric sample have peak absolute magnitudes either around or below the spectroscopic follow-up depth, which explains why these event were not chosen for spectroscopic follow-up. However, there are 5 photometrically-classified SLSNe (PSc061198, PSc080492, PSc110446, PSc390605, and PSc490019) at lower redshift ($z\le 0.6$) that are more than 1 magnitude brighter than the threshold but were not chosen as follow-up candidates. 

\subsection{Physical Properties and Correlations}
\label{sec:phys_prop}

\begin{figure*}
    \centering
    \includegraphics[width=\textwidth]{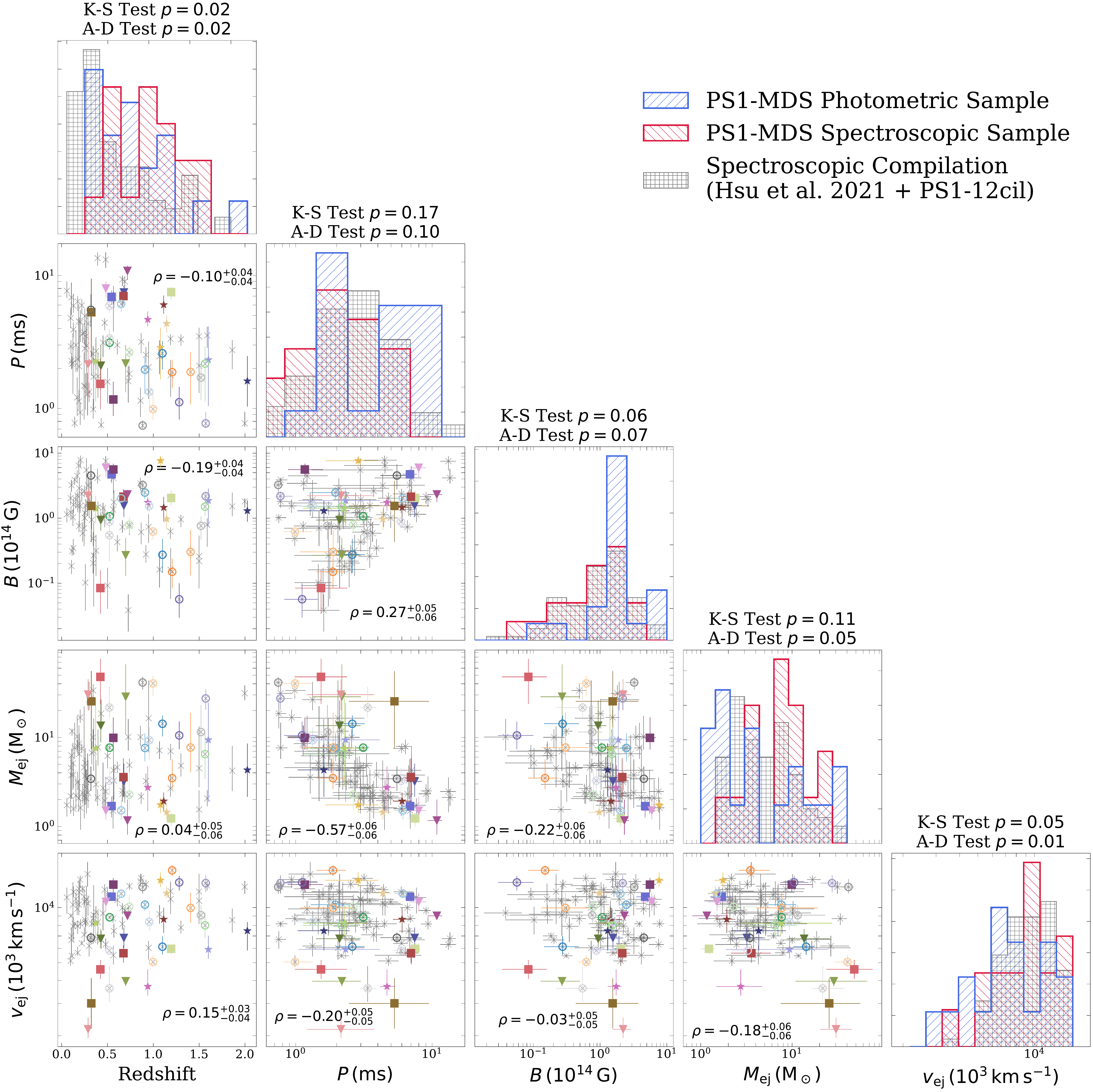}    
    \caption{Median values and 1$\sigma$ uncertainties of the key magnetar model parameters ($P$, $B$, $M_{\rm ej}$, $\varv_{\rm ej}$) (solid squares: photometric; open circles: spectroscopic; plot symbols for the photometrically-classified SLSNe are the same as in Figure~\ref{fig:mag_z}). The models for PS1-11ap and PS1-12cil are both obtained from \citet{Hosseinzadeh_2021}. The gray crosses mark the remaining spectroscopically confirmed SLSNe from \citet{Hsu_2021}. In the top panels we show the parameter distributions for the PS1-MDS photometric sample (blue), PS1-MDS spectrosc sample (red), and the SLSN compilation sample (grey), along with the median $p$-values associated with both the K-S test and the A-D test statistics, calculated using the PS1-MDS photometric sample and the SLSN compilation. In each panel we quote the median value and 1$\sigma$ bound of the Spearman rank correlation coefficient using the PS1-MDS photometric sample and literature data set. Of all parameter pairs, $P$ and $M_{\rm ej}$ exhibit the strongest correlation, consistent with the findings in \citet{Blanchard_2020} and \citet{Hsu_2021}.}
    \label{fig:phys_params}
\end{figure*}

In Figure~\ref{fig:phys_params} we show two-dimensional distributions of the primary physical parameters ($P$, $B$, $M_{\rm ej}$, and $\varv_{\rm ej}$; the medians of the posteriors) and redshifts of both PS1-MDS samples and the SLSN compilation, which contains events from a wide range of surveys (including the PS1-MDS spectroscopic sample). We explore both differences between the three samples, and parameter correlations for the combined sample (all three samples together, 101 SLSNe in total). Specifically, we compare the PS1-MDS photometric sample and the spectroscopic compilation sample using the two-sample Kolmogorov-Smirnov (K-S) test \citep{KS_test} and the two-sample Anderson-Darling (A-D) test \citep{AD_test}. Both tests are designed to determine whether two distributions arise from the same underlying population. The A-D test is a modification of the K-S test that is more sensitive to the tails of a distribution, whereas the K-S test gives more weight to the mean of a distribution. We report the resulting $p$-values from these tests, to determine if both are drawn from the same parameter distribution, at the top of each column in Figure~\ref{fig:phys_params}. 

The differences in redshift distributions between the two samples reflect the design characteristics of the various surveys (e.g., PS1-MDS, Dark Energy Survey, PTF, etc). In terms of the magnetar model parameters we find that the distributions are overall in good agreement, except for the ejecta velocity, which has statistically significant $p$-values for the A-D test. This indicates that we can reject the null hypothesis at 95\% conficence that the ejecta velocity for the photometric and the spectroscopic compilation samples are drawn from the same distribution. This may be caused by the sensitivity of the A-D test to tail distributions. The spectroscopic compilation sample spans a range of $\varv_{\rm ej}\approx (3.6-16)\times10^3$ km s$^{-1}$, while the photometric sample spans a range of $\varv_{\rm ej}\approx (2.2-14)\times10^3$ km s$^{-1}$, with two events\footnote{The two $\varv_{\rm ej}$ outliers (PSc130096 and PSc390605) have relatively few data points. PSc130096 lacks a definitive peak and any post-peak data, and the model is therefore only marginally constrained. PSc390605 similarly lacks pre- and post-peak again leading to a marginally constrained model.} (PSc130096 and PSc390605) having $\varv_{\rm ej}$ values that fall outside the range of the spectroscopic population. Removing these two outliers return an updated A-D test $p$-value of $\approx 0.06$, suggesting that other than these two specific data points, the remainder of the photometric sample fit into the spectroscopic sample well.  We explore the posterior distributions of the magnetar parameters from the photometric sample in more detail in the next subsection.

As done in previous SLSN parameter studies (e.g., \citealt{Blanchard_2020,Hsu_2021}), we combine the PS1-MDS photometric and literature samples to confirm known correlations and explore new ones. For each pair of parameters, we perform a Monte Carlo procedure to calculate the Spearman rank correlation coefficient \citep[$\rho$;][]{Spearman_1904} and its associated $1\sigma$ bound using the method described in \citet{Curran_2014}. The results are summarized in each panel of Figure~\ref{fig:phys_params}. We find the same results as \citet{Hsu_2021}, where most parameter combinations exhibit either no correlation, mild  correlations, or mild correlations that are primarily due to the absence of events in specific areas of the parameter space. The mass-spin correlation discussed first in \citet{Blanchard_2020} remains strong after merging the photometric and spectroscopic samples. All other mild correlations have been previously explained as being due to observational biases in \citet{Blanchard_2020} and \citet{Hsu_2021}, and we do not find any new statistically significant correlations here.

\subsection{Posterior Distributions of the Photometric Sample}
\label{sec:phot_posteriors}

\begin{figure*}[t!]
    \centering
    \includegraphics[width=\textwidth]{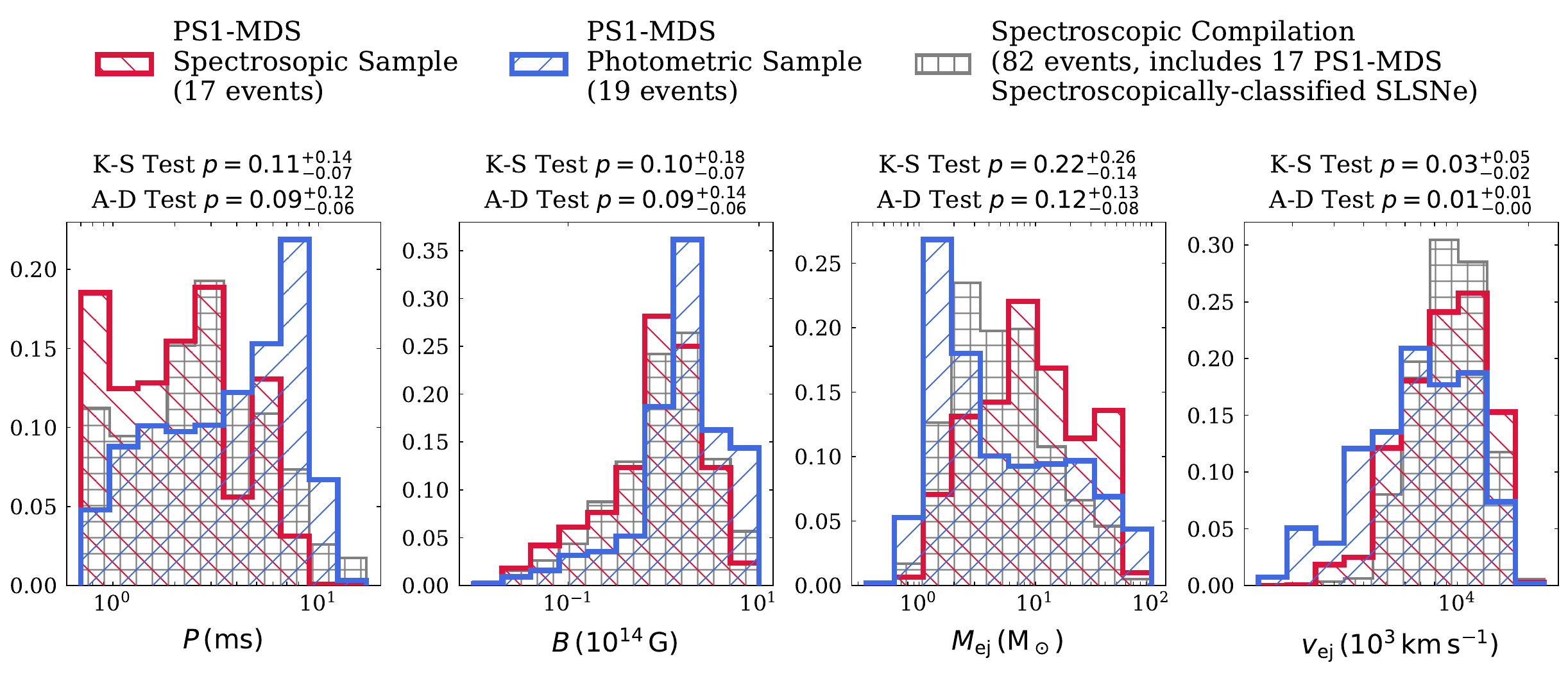}
    \caption{Normalized joint posterior distributions of the magnetar and ejecta  parameters for the PS1-MDS photometric (blue), PS1-MDS spectroscopic (red), and the spectroscopic compilation (grey) samples. The distributions are summed posteriors of the {\tt MOSFiT} parameters, consisting of 100 randomly sampled MCMC walkers from each SLSN fit. At the top of each panel we quote the median $p$-values and their 1$\sigma$ ranges from both the K-S and the A-D tests using the bootstrap method by comparing the PS1-MDS photometric sample and the SLSN compilation.}
    \label{fig:joint_posterior}
\end{figure*}

To explore any differences in magnetar and ejecta parameters between the PS1-MDS photometric and spectroscopic samples, we show in Figure~\ref{fig:joint_posterior} the joint posterior distributions of the PS1-MDS photometric, PS1-MDS spectroscopic, and the compilation samples. We construct the joint posterior distributions by selecting 100 randomly sampled walkers from each \texttt{MOSFiT} fit.\footnote{We take 100 here instead of the full 120 walkers as described in \S\ref{sec:fits} because some events modeled previously in \citet{Blanchard_2020} only have 100 walkers.}

To capture uncertainties in the test statistics, we calculate and report in each panel the two-sample K-S test and the two-sample A-D test $p$-values between the PS1-MDS photometric sample and the spectroscopic compilation using a modified bootstrap method. For each parameter, we calculate a distribution of $p$-values by repeating the following procedure 5000 times. We assemble a joint posterior for the 19 photometrically-classified SLSNe by randomly drawing one MCMC walker from the individual posterior for each event, and we do the same for the 82 spectroscopically-classified SLSNe. We then calculate $p$-values for the K-S and A-D tests comparing these two joint posteriors. We report the median and $1\sigma$ bounds of these distributions of the resulting $p$-values on top of each panel in Fig~\ref{fig:joint_posterior}. 

The posterior distributions for the physical parameters are in good agreement, except for $\varv_{\rm ej}$, as noted previously; removing PSc130096 and PSc390605 from the photometric sample leads to $p={0.11}_{-0.08}^{+0.15}$ (K-S) and $p={0.05}_{-0.03}^{+0.06}$ (A-D). We also note that while the K-S and A-D tests indicate that the distributions of $P$ and $M_{\rm ej}$ are drawn from the same distribution, the photometric sample skews to slower spins and lower ejecta masses (this trend is still in agreement with the mass-spin correlation).  This difference can be ascribed to the systematically lower luminosities of the photometric SLSNe (Figure~\ref{fig:mag_z}) compared to the PS1-MDS spectroscopic SLSNe.

\subsection{Effects of Classification Uncertainty}

\begin{figure*}
    \centering
    \includegraphics[width=\textwidth]{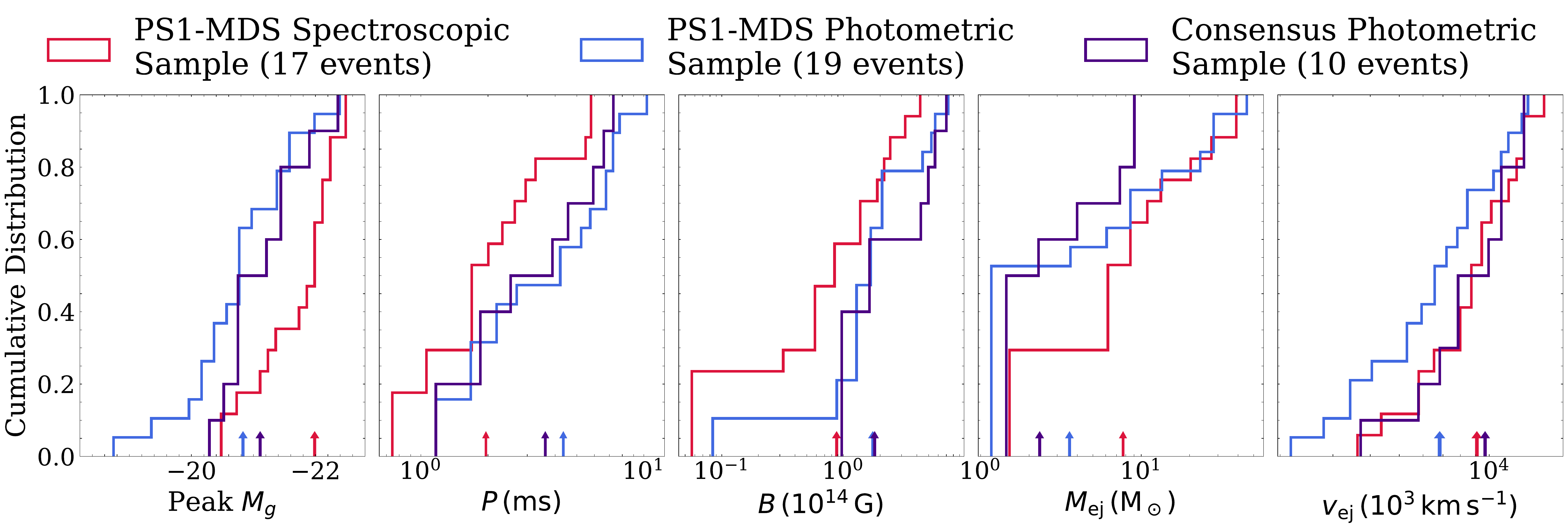}
    \caption{Cumulative distributions of peak absolute rest-frame $g$-band magnitude and median magnetar model parameter values for the PS1-MDS spectroscopic (red), photometric (blue), and consensus photometric (purple) samples. The arrows in each panel indicate the median parameter value for the samples. The consensus photometric sample contains only events classified as SLSNe by both classifiers. It exhibits a higher median magnitude of $\approx-21.1$ (purple arrow) compared to $\approx-20.8$ for the full photometric sample, but is still $\approx 1$ magnitude dimmer than the median of $\approx-22$ for the spectroscopic sample. Even though the full and the consensus photometric samples have comparable median $B$ values ($\approx 1.74\times10^{15}$ G and $1.80\times10^{15}$ G for the full and consensus samples, respectively), the consensus sample spans a much narrower and higher range in $B$. The shift in $M_{\rm ej}$ is more strongly reflected, with a lower median value ($\approx 3.58$ M$_{\odot}$, full; $\approx 2.33$ M$_{\odot}$, consensus) at a lower range. All of the shifts in magentar model parameters are consistent with SLSNe with higher luminosities than the full photometric sample.}
    \label{fig:med_comp}
\end{figure*}

As indicated in Table~\ref{tab:classifications}, 9 of the 19 photometrically-classified SLSNe in our final sample were designated as SLSNe by only one of the two classifiers. To investigate the impact of these cases of classification disagreement, we repeat the analyses in the previous subsections using only events classified as SLSNe by both {\tt Superphot} and {\tt SuperRAENN}. This ``consensus'' photometric sample spans a peak absolute magnitude range of $\approx -20.3$ to $\approx -22.6$. However, despite excluding some of the lowest luminosity events, the median peak magnitude is still $\approx 1$ mag dimmer than that of the spectroscopic sample (see Figure~\ref{fig:med_comp}, left), and we find the same trend of systematically lower luminosity at any redshift as seen for the full sample in Figure~\ref{fig:mag_z}. Our conclusion about the lower luminosities probed by the photometric sample thus remains unchanged.

Systematically removing objects classified as SLSNe by only one classifier eliminates the disagreement in the $\varv_{\rm ej}$ distributions but introduces mildly statistically significant differences in $B$ and $M_{\rm ej}$. The consensus sample shifts to higher ranges of $B\approx(1-7.7)\times10^{14}$ G, $\varv_{\rm ej}\approx(0.37-1.41)\times10^4)$ km s$^{-1}$, a lower range of $M_{\rm ej}\approx 1.4-9.9$ M$_\odot$, and a similar range of $P\approx(1.17-7.98)$ ms in parameter distributions. These shifts are all consistent and expected for SLSNe with higher luminosities. See Figure~\ref{fig:med_comp} for these changes in magnetar model parameters. The shift in $B$ is reflected in the posterior distribution but not as strongly in $M_{\rm ej}$.

\section{Discussion and Conclusions}
\label{sec:conclusions}

In this paper we presented a case study for time-domain science with machine learning-based photometric classification, focusing on SLSNe from the PS1-MDS.  Our analysis consisted of two critical aspects that would need to be undertaken for any future such studies (for SLSNe or any other types of transients).  First, we began with a sample of events nominally classified as SLSNe by two independent machine learning-based pipelines (\texttt{SuperRAENN} and \texttt{SuperPhot}).  We then applied various selection criteria to increase the sample purity (e.g., removing likely AGN flares, setting a higher minimum classification probability threshold) at the cost of sample completeness.  Our sample size following these cuts was 36\% of the initial sample (24 of the 67). Subsequent to the sample refinement we carried out modeling with \texttt{MOSFiT} to extract physical parameters in order to compare the photometric sample with existing spectroscopic samples modeled in the same way.  The requirement for model convergence eliminated 5 additional events from the sample (21\% reduction from 24 to 19).  These two critical steps of sample refinement and modeling will be essential for all studies with photometrically-classified samples.

Comparing our photometric SLSN sample to the PS1-MDS spectroscopically-classified SLSNe and to the larger sample of spectroscopic SLSNe, we find an overall similarity in both observed properties and inferred magnetar and ejecta parameters. We do note a potential shift in the photometric sample to slower magnetar spins and lower ejecta masses, which may reflect the fact that the photometric SLSNe are systematically dimmer than the spectroscopic PS1-MDS SLSNe (due to the shallower effective magnitude limit required for spectroscopy).  If this is indeed the case, then it highlights an important advantage of photometric classification in deep surveys (such as PS1-MDS and LSST). 

Our initial classifications and the subsequent modeling both rely on the existence of redshift information.  In the case of our PS1-MDS sample, the redshifts were determined from host galaxy spectroscopy after the survey concluded.  Such data may be difficult to obtain for the large samples expected from LSST (e.g., $\gtrsim 10^6$ SNe per year, and $\sim 10^4$ SLSNe per year \citealt{Villar_2018}). However, robust photometric redshifts are likely to be as useful as spectroscopic redshifts.  We also note that one source of contamination in our initial photometric sample appears to be AGN (21\%, 14 of 67 events) despite the fact that the PS1-MDS sample was designed to eliminate variable AGN. These contaminating AGN were again identified via host galaxy spectroscopy, which will not be available for the LSST samples; a more robust elimination of AGN will be essential.  

Overall, our analysis highlights some challenges in constructing pure samples of photometrically-classified SNe, but we believe that these challenges are surmountable.  The photometric sample explored here is smaller than the overall known spectroscopic sample by a factor of several, but looking forward to LSST, even a highly conservative selection with relatively low completeness will easily exceed the spectroscopic sample by two orders of magnitude.   

\vspace{12pt}  
The Berger Time Domain group at Harvard is supported in part by NSF and NASA grants, including support by the NSF under grant AST-2108531, as well as by the NSF under Cooperative Agreement PHY-2019786 (The NSF
AI Institute for Artificial Intelligence and Fundamental Interactions http://iafi.org/). VAV acknowledges support in part by the NSF through grant AST-2108676. The computations presented in this work were performed on the FASRC Cannon cluster supported by the FAS Division of Science Research Computing Group at Harvard University.

\vspace{12pt} \noindent \textit{Software}: Astropy \citep{Astropy_2013,Astropy_2018}, extinction \citep{Barbary_2016}, \texttt{MOSFiT} \citep{Guillochon_2018}, Matplotlib \citep{Hunter_2007}, NumPy \citep{Oliphant_2006}, pymccorrelation \citep{Curran_2014,Privon_2020}, Scipy \citep{Virtanen_2020}, {\tt Superphot} \citep{Hosseinzadeh_2020}, {\tt SuperRAENN} \citep{Villar_2020}

\noindent \facility{ADS, PS1}

\bibliography{Reference}
\bibliographystyle{aasjournal}
\end{document}

%% file: affil.tex
\newcommand{\CfA}{\affiliation{Center for Astrophysics \textbar{} Harvard \& Smithsonian, 60 Garden Street, Cambridge, MA 02138-1516, USA}}
\newcommand{\UA}{\affiliation{Steward Observatory, University of Arizona, 933 North Cherry Avenue, Tucson, AZ 85721-0065, USA}}

\newcommand{\IAIFI}{\affiliation{The NSF AI Institute for Artificial Intelligence and Fundamental Interactions}}

%% file: cut_seq.tex
\begin{deluxetable*}{lcccc}
\tablewidth{\columnwidth} 
\caption{Sequential Selection Criteria\label{tab:seq}}
\tablehead{\colhead{Metric} & 
\colhead{} &
\colhead{} &
\colhead{Both} & \colhead{Total SLSNe}\\[-12pt]
\colhead{} & 
\colhead{SuperRAENN} & \colhead{Superphot} & \colhead{} & \colhead{}\\[-12pt]
\colhead{(Applied Sequentially)} & 
\colhead{} &
\colhead{} & \colhead{Algorithms} &
\colhead{Classified}\tablenotemark{$\ast$}}
\startdata
No criteria applied & 37 & 58 & 28 & 67 \\
Not within $1''$ of AGN host center & 25 & 44 & 16 & 53 \\
Classification confidence $\ge$ 0.5 & 18 & 28 & 10 & 36 \\
$\ge$ 11  detection across all 4 bands & 16 & 17 & 9 & 24 \\
PSRF $\le 1.1$ & 13 & 13 & 7 & 19 \\[+2pt]
\enddata
\tablenotetext{\ast}{The total number of photometrically-classified SLSNe takes into account events classified by both algorithms.}
\end{deluxetable*}

%% file: classifications.tex
\begin{deluxetable}{lcccc}
\tablewidth{\columnwidth} 
\caption{Classification Results for Final SLSN Photometric Sample}
\tablehead{\colhead{} & \multicolumn{2}{c}{{\tt SuperRAENN}} & \multicolumn{2}{c}{{\tt Superphot}} \\[-10pt]
\colhead{PScID} & \multicolumn{2}{c}{------------------------------} & \multicolumn{2}{c}{------------------------------}\\[-10pt]
\colhead{} & \colhead{SN Type} & \colhead{Confidence} & \colhead{SN Type} & \colhead{Confidence}}
\startdata
PSc000036 & SLSN & 1.00 & SLSN & 0.89\\
PSc000553\tablenotemark{$\dagger$} & SLSN & 0.84 & SNIIn & 0.52\\
PSc061198 & SLSN & 0.41 & SLSN & 0.82\\
PSc070299 & SLSN & 1.00 & SLSN & 0.99\\
PSc080492\tablenotemark{$\dagger$} & SLSN & 0.86 & SNIIn & 0.50\\
PSc091753\tablenotemark{$\dagger$} & SLSN & 0.78 & SNIIn & 0.47\\
PSc110446 & SLSN & 0.94 & SLSN & 0.94\\
PSc120151\tablenotemark{$\dagger$} & SNIIn & 0.66 & SLSN & 0.86\\
PSc130096\tablenotemark{$\dagger$} & SNIa & 0.93 & SLSN & 0.71\\
PSc300035 & SLSN & 0.76 & SLSN & 0.89\\
PSc310006 & SLSN & 1.00 & SLSN & 0.98\\
PSc320338 & SLSN & 0.98 & SLSN & 0.94\\
PSc380044\tablenotemark{$\dagger$} & SNIIn & 0.39 & SLSN & 0.58\\
PSc390461\tablenotemark{$\dagger$} & SNIa & 1.00 & SLSN & 0.77\\
PSc390605\tablenotemark{$\dagger$} & SLSN & 0.64 & SNIIn & 0.95\\
PSc420350 & SLSN & 0.39 & SLSN & 0.69\\
PSc450057\tablenotemark{$\dagger$} & SLSN & 0.51 & SNIIn & 0.61\\
PSc480628 & SLSN & 0.61 & SLSN & 0.73\\
PSc490019 & SLSN & 0.60 & SLSN & 0.39\\[+2pt]
\enddata
\tablenotetext{\dagger}{Event classified as a SLSN by only one classifier.}
\tablecomments{Classification results for the final 19 photometric SLSNe from both {\tt SuperRAENN} and {\tt Superphot}. Here, we adopt the SN class with the highest classification probability as the predicted SN type for each transient. If either algorithm classifies an event as a SLSN, we include it in our sample.}
\label{tab:classifications}
\end{deluxetable}

%% file: priors.tex
\begin{deluxetable}{lccccc}
\tablewidth{\columnwidth} 
\caption{Priors on the Magnetar Model Parameters}
\tablehead{\colhead{Parameter} & 
\colhead{Prior} & 
\colhead{Lower} & 
\colhead{Upper} & 
\twocolhead{Gaussian} \\[-6pt]
\colhead{/Units} &
\colhead{Shape} &
\colhead{Bound} &
\colhead{Bound} &
\colhead{Mean} & 
\colhead{S.D.}}
\startdata
$P/\rm ms$ & Flat & 0.7 & 20 & $\cdots$ & $\cdots$\\
$B/10^{14}\ \rm G$ & Flat & 0.1 & 10 & $\cdots$ & $\cdots$\\
$M_{\rm ej}$/$M_{\odot}$ & Flat & 0.1 & 100 & $\cdots$ & $\cdots$\\
$\varv_{\rm ej}/10^4\ \rm km\ s^{-1}$ & Gaussian & 0.1 & 3.0 & 1.47 & 4.3\\
$\kappa/\ \rm g\ cm^{-2}$ & Flat  & 0.05 & 0.2 & $\cdots$ & $\cdots$\\
$\kappa_{\gamma}/\ \rm g\ cm^{-2}$ & Log-flat  & 0.01 & 100 & $\cdots$ & $\cdots$\\
$M_{\rm NS}$/$M_{\odot}$ & Flat  & 1.4 & 2.2 & $\cdots$ & $\cdots$\\
$T_{\rm min}/10^3\ \rm K$ & Gaussian & 3.0 & 10.0 & 6.0 & 1.0\\
$n_{\rm H,host}$/cm$^{-2}$& Log-flat  & $10^{16}$ & $10^{23}$ & $\cdots$ & $\cdots$\\
$t_{\rm exp}/\rm days$ & Flat  & $-100$ & 0 & $\cdots$ & $\cdots$\\
$\sigma$/mag & Log-flat  & 10$^{-3}$ & 100 & $\cdots$ & $\cdots$\\[+2pt]
\enddata
\tablecomments{$P$ is the initial spin period of the magnetar; $B$ is the magnetic field strength; $M_{\rm ej}$ is the ejecta mass; $\varv_{\rm ej}$ is the ejecta velocity; $\kappa$ is the opacity; $\kappa_{\gamma}$ is the gamma-ray opacity; $M_{\rm NS}$ is the neutron star mass; $T_{\rm min}$ is the photospheric temperature floor; $n_{\rm H,host}$ is the hydrogen column density in the host galaxy, a proxy for extinction; $t_{\rm exp}$ is the time of explosion relative to the first observed data point; $\sigma$ is the additional photometric uncertainty required to yield a reduced chi-squared value of ${\approx}1$. All priors, including Gaussian priors, are bounded as specified above. For a detailed description of the model see \citet{Nicholl_2017b}.}
\label{tab:priors}
\end{deluxetable}

%% file: phys_params.tex
\begin{deluxetable*}{lcccc}
\tablewidth{\textwidth} 
\caption{\added{Median} Magnetar Parameters Values for the Photometrically-Classified SLSNe}
\tablehead{\colhead{} & \colhead{$P$} & \colhead{$B$} & \colhead{$M_{\rm ej}$} &
\colhead{$\varv_{\rm ej}$}\\[-13pt]
\colhead{SLSN Name} & \colhead{} & \colhead{} & \colhead{} & \colhead{} \\[-12pt]
\colhead{} & 
\colhead{(ms)} & 
\colhead{(10$^{14}$ G)} &
\colhead{(M$_\odot$)} &
\colhead{(10$^3$ km s$^{-1}$)}}

\startdata
PSc000036 & $\phn{1.61}_{-0.58}^{+0.87}$ & ${1.30}_{-0.42}^{+0.57}$ & $\phn{4.31}_{-2.16}^{+4.25}$ & $\phn{7.47}_{-1.64}^{+0.85}$ \\
PSc000553\tablenotemark{$\dagger$} & $\phn{7.45}_{-0.95}^{+0.97}$ & ${1.56}_{-0.51}^{+0.67}$ & $\phn{3.19}_{-1.04}^{+1.87}$ & $\phn{6.83}_{-0.80}^{+0.96}$ \\
PSc061198 & $\phn{6.89}_{-0.84}^{+0.82}$ & ${4.73}_{-1.25}^{+0.91}$ & $\phn{1.69}_{-0.44}^{+0.46}$ & ${11.46}_{-0.53}^{+0.46}$ \\
PSc070299 & $\phn{2.32}_{-1.28}^{+1.81}$ & ${1.87}_{-1.13}^{+1.00}$ & $\phn\phn{9.38}_{-7.50}^{+15.07}$ & $\phn{5.91}_{-0.65}^{+1.12}$ \\
PSc080492\tablenotemark{$\dagger$} & $\phn{2.10}_{-0.69}^{+0.90}$ & ${0.94}_{-0.29}^{+0.30}$ & ${13.54}_{-4.75}^{+12.43}$ & $\phn{6.76}_{-0.74}^{+0.87}$ \\
PSc091753\tablenotemark{$\dagger$} & $\phn{2.18}_{-0.86}^{+0.83}$ & ${0.27}_{-0.14}^{+0.23}$ & $\phn{28.62}_{-17.73}^{+37.64}$ & $\phn{3.97}_{-0.24}^{+0.21}$ \\
PSc110446 & $\phn{2.27}_{-1.14}^{+1.07}$ & ${1.40}_{-0.54}^{+0.78}$ & $\phn{7.52}_{-4.74}^{+19.93}$ & $\phn{8.22}_{-1.22}^{+1.98}$ \\
PSc120151\tablenotemark{$\dagger$} & $\phn{7.51}_{-0.93}^{+0.69}$ & ${2.04}_{-0.49}^{+0.64}$ & $\phn{1.23}_{-0.23}^{+0.20}$ & $\phn{5.96}_{-0.38}^{+0.30}$ \\
PSc130096\tablenotemark{$\dagger$} & $\phn{5.30}_{-2.84}^{+4.18}$ & ${1.55}_{-1.12}^{+2.20}$ & ${25.45}_{-17.95}^{+30.36}$ & $\phn{3.00}_{-0.82}^{+1.12}$ \\
PSc300035 & $\phn{2.87}_{-1.22}^{+1.16}$ & ${7.66}_{-1.55}^{+1.36}$ & $\phn{1.75}_{-0.33}^{+0.33}$ & ${14.09}_{-1.16}^{+1.47}$ \\
PSc310006 & $\phn{4.35}_{-0.62}^{+0.66}$ & ${0.97}_{-0.31}^{+0.41}$ & $\phn{1.44}_{-0.25}^{+0.38}$ & ${11.99}_{-0.79}^{+1.08}$ \\
PSc320338 & $\phn{6.02}_{-0.86}^{+1.06}$ & ${1.46}_{-0.56}^{+0.67}$ & $\phn{1.93}_{-0.46}^{+0.47}$ & $\phn{8.63}_{-1.36}^{+1.36}$ \\
PSc380044\tablenotemark{$\dagger$} & $\phn{7.00}_{-0.95}^{+0.88}$ & ${2.14}_{-0.52}^{+0.65}$ & $\phn{3.58}_{-1.44}^{+2.25}$ & $\phn{5.64}_{-0.64}^{+0.58}$ \\
PSc390461\tablenotemark{$\dagger$} & $\phn{1.53}_{-0.55}^{+0.85}$ & ${0.08}_{-0.05}^{+0.07}$ & ${47.89}_{-24.13}^{+28.57}$ & $\phn{4.60}_{-0.53}^{+0.57}$ \\
PSc390605\tablenotemark{$\dagger$} & $\phn{2.15}_{-0.94}^{+1.61}$ & ${2.21}_{-0.67}^{+0.68}$ & ${30.24}_{-10.10}^{+14.93}$ & $\phn{2.17}_{-0.22}^{+0.34}$ \\
PSc420350 & $\phn{1.17}_{-0.29}^{+0.42}$ & ${5.59}_{-1.27}^{+1.30}$ & $\phn{9.90}_{-2.60}^{+6.85}$ & ${13.36}_{-1.26}^{+1.14}$ \\
PSc450057\tablenotemark{$\dagger$} & ${10.83}_{-1.56}^{+1.03}$ & ${2.31}_{-0.56}^{+0.42}$ & $\phn{1.16}_{-0.36}^{+0.84}$ & $\phn{9.02}_{-0.17}^{+0.26}$ \\
PSc480628 & $\phn{4.67}_{-0.52}^{+0.71}$ & ${1.74}_{-0.44}^{+0.36}$ & $\phn{2.74}_{-1.25}^{+1.97}$ & $\phn{3.71}_{-0.40}^{+0.89}$ \\
PSc490019 & $\phn{7.98}_{-1.15}^{+1.07}$ & ${5.95}_{-1.72}^{+0.91}$ & $\phn{1.50}_{-0.30}^{+0.65}$ & ${10.74}_{-1.17}^{+1.34}$ \\[+2pt]
\enddata
\tablenotetext{\dagger}{Event classified as a SLSN by only one classifier.}
\label{tab:phys_params}
\end{deluxetable*}

%% file: median.tex
\begin{deluxetable}{lccc}[t!]
\tablewidth{\columnwidth} 
\caption{Magnetar Parameter Sample Median Values}
\tablehead{\colhead{} & \colhead{PS1-MDS} & \colhead{PS1-MDS} & \colhead{SLSN}\\[-12pt]
\colhead{Parameter} & \colhead{} & \colhead{} & \colhead{}\\[-12pt]
\colhead{} & \colhead{Photometric} & \colhead{Spectroscopic} & \colhead{Compilation}}
\startdata
$P$ (ms)& $4.35_{-2.32}^{+3.11}$ & $1.96_{-0.90}^{+2.36}$ & $2.67_{- 1.34}^{+3.26}$\\
$B$ ($10^{14}$ G) & $1.56_{-0.63}^{+3.28}$ & $0.88_{-0.59}^{+1.43}$ & $1.14_{- 0.86}^{+1.75}$\\
$M_{\rm ej}$ (M$_{\odot}$) & $3.19_{-1.70}^{+22.6}$ & $7.70_{- 4.35}^{+16.5}$ & $4.56_{-2.37}^{+6.02}$\\
$\varv_{\rm ej}$ ($10^3$ km s) & $6.83_{-2.89}^{+4.69}$ & $9.09_{-3.10}^{+4.15}$ & $9.25_{-2.58}^{+3.40}$\\
$E_K$ ($10^{51}$ erg) & $2.21_{-1.10}^{+2.98}$ & $6.48_{-4.24}^{+7.93}$ & $3.78_{-1.86}^{+5.09}$\\[+2pt]
\enddata
\tablecomments{The median values and $1\sigma$ ranges for the magnetar engine and ejecta parameters of the PS1-MDS SLSN samples (photometric and spectroscopic), and the SLSN compilation sample (from \citealt{Hsu_2021}, with the addition of PS1-12cil), which include the 17 PS1-MDS spectroscopically-classified SLSNe.}
\label{tab:median}
\end{deluxetable}

%% file: main.bbl
\begin{thebibliography}{}
\footnotesize
\expandafter\ifx\csname natexlab\endcsname\relax\def\natexlab#1{#1}\fi
\providecommand{\url}[1]{\href{#1}{#1}}
\providecommand{\dodoi}[1]{doi:~\href{http://doi.org/#1}{\nolinkurl{#1}}}

\bibitem[{{Anderson} \& {Darling}(1952)}]{AD_test}
{Anderson}, T.~W., \& {Darling}, D.~A. 1952,
  \hypersetup{urlcolor=magenta}\href{https://dx.doi.org/10.1214/aoms/1177729437}{The
  Annals of Mathematical Statistics}, 23, 193

\bibitem[{{Astropy Collaboration} {et~al.}(2013){Astropy Collaboration},
  {Robitaille}, {Tollerud}, {Greenfield}, {Droettboom}, {Bray}, {Aldcroft},
  {Davis}, {Ginsburg}, {Price-Whelan}, {Kerzendorf}, {Conley}, {Crighton},
  {Barbary}, {Muna}, {Ferguson}, {Grollier}, {Parikh}, {Nair}, {Unther},
  {Deil}, {Woillez}, {Conseil}, {Kramer}, {Turner}, {Singer}, {Fox}, {Weaver},
  {Zabalza}, {Edwards}, {Azalee Bostroem}, {Burke}, {Casey}, {Crawford},
  {Dencheva}, {Ely}, {Jenness}, {Labrie}, {Lim}, {Pierfederici}, {Pontzen},
  {Ptak}, {Refsdal}, {Servillat}, \& {Streicher}}]{Astropy_2013}
{Astropy Collaboration}, {Robitaille}, T.~P., {Tollerud}, E.~J., {et~al.} 2013,
  \hypersetup{urlcolor=magenta}\href{https://dx.doi.org/10.1051/0004-6361/201322068}{A\&A},
  \hypersetup{urlcolor=blue}\href{https://ui.adsabs.harvard.edu/abs/2013A&A...558A..33A}{558,
  A33}

\bibitem[{{Astropy Collaboration} {et~al.}(2018){Astropy Collaboration},
  {Price-Whelan}, {Sip{\H{o}}cz}, {G{\"u}nther}, {Lim}, {Crawford}, {Conseil},
  {Shupe}, {Craig}, {Dencheva}, {Ginsburg}, {Vand erPlas}, {Bradley},
  {P{\'e}rez-Su{\'a}rez}, {de Val-Borro}, {Aldcroft}, {Cruz}, {Robitaille},
  {Tollerud}, {Ardelean}, {Babej}, {Bach}, {Bachetti}, {Bakanov}, {Bamford},
  {Barentsen}, {Barmby}, {Baumbach}, {Berry}, {Biscani}, {Boquien}, {Bostroem},
  {Bouma}, {Brammer}, {Bray}, {Breytenbach}, {Buddelmeijer}, {Burke},
  {Calderone}, {Cano Rodr{\'\i}guez}, {Cara}, {Cardoso}, {Cheedella}, {Copin},
  {Corrales}, {Crichton}, {D'Avella}, {Deil}, {Depagne}, {Dietrich}, {Donath},
  {Droettboom}, {Earl}, {Erben}, {Fabbro}, {Ferreira}, {Finethy}, {Fox},
  {Garrison}, {Gibbons}, {Goldstein}, {Gommers}, {Greco}, {Greenfield},
  {Groener}, {Grollier}, {Hagen}, {Hirst}, {Homeier}, {Horton}, {Hosseinzadeh},
  {Hu}, {Hunkeler}, {Ivezi{\'c}}, {Jain}, {Jenness}, {Kanarek}, {Kendrew},
  {Kern}, {Kerzendorf}, {Khvalko}, {King}, {Kirkby}, {Kulkarni}, {Kumar},
  {Lee}, {Lenz}, {Littlefair}, {Ma}, {Macleod}, {Mastropietro}, {McCully},
  {Montagnac}, {Morris}, {Mueller}, {Mumford}, {Muna}, {Murphy}, {Nelson},
  {Nguyen}, {Ninan}, {N{\"o}the}, {Ogaz}, {Oh}, {Parejko}, {Parley}, {Pascual},
  {Patil}, {Patil}, {Plunkett}, {Prochaska}, {Rastogi}, {Reddy Janga},
  {Sabater}, {Sakurikar}, {Seifert}, {Sherbert}, {Sherwood-Taylor}, {Shih},
  {Sick}, {Silbiger}, {Singanamalla}, {Singer}, {Sladen}, {Sooley},
  {Sornarajah}, {Streicher}, {Teuben}, {Thomas}, {Tremblay}, {Turner},
  {Terr{\'o}n}, {van Kerkwijk}, {de la Vega}, {Watkins}, {Weaver}, {Whitmore},
  {Woillez}, {Zabalza}, \& {Astropy Contributors}}]{Astropy_2018}
{Astropy Collaboration}, {Price-Whelan}, A.~M., {Sip{\H{o}}cz}, B.~M., {et~al.}
  2018,
  \hypersetup{urlcolor=magenta}\href{https://dx.doi.org/10.3847/1538-3881/aabc4f}{AJ},
  \hypersetup{urlcolor=blue}\href{https://ui.adsabs.harvard.edu/abs/2018AJ....156..123A}{156,
  123}

\bibitem[{Barbary(2016)}]{Barbary_2016}
Barbary, K. 2016,
  \hypersetup{urlcolor=magenta}\href{https://dx.doi.org/10.21105/joss.00058}{JOSS},
  \hypersetup{urlcolor=blue}\href{https://ui.adsabs.harvard.edu/abs/2016JOSS....1...58B}{1,
  58}

\bibitem[{{Bellm} {et~al.}(2019){Bellm}, {Kulkarni}, {Barlow}, {Feindt},
  {Graham}, {Goobar}, {Kupfer}, {Ngeow}, {Nugent}, {Ofek}, {Prince}, {Riddle},
  {Walters}, \& {Ye}}]{Bellm_2019}
{Bellm}, E.~C., {Kulkarni}, S.~R., {Barlow}, T., {et~al.} 2019,
  \hypersetup{urlcolor=magenta}\href{https://dx.doi.org/10.1088/1538-3873/ab0c2a}{\pasp},
  \hypersetup{urlcolor=blue}\href{https://ui.adsabs.harvard.edu/abs/2019PASP..131f8003B}{131,
  068003}

\bibitem[{{Blanchard} {et~al.}(2021){Blanchard}, {Berger}, {Nicholl},
  {Chornock}, {Gomez}, \& {Hosseinzadeh}}]{Blanchard_2021}
{Blanchard}, P.~K., {Berger}, E., {Nicholl}, M., {et~al.} 2021,
  \hypersetup{urlcolor=magenta}\href{https://dx.doi.org/10.3847/1538-4357/ac1b27}{\apj},
  \hypersetup{urlcolor=blue}\href{https://ui.adsabs.harvard.edu/abs/2021ApJ...921...64B}{921,
  64}

\bibitem[{{Blanchard} {et~al.}(2020){Blanchard}, {Berger}, {Nicholl}, \&
  {Villar}}]{Blanchard_2020}
{Blanchard}, P.~K., {Berger}, E., {Nicholl}, M., \& {Villar}, V.~A. 2020,
  \hypersetup{urlcolor=magenta}\href{https://dx.doi.org/10.3847/1538-4357/ab9638}{ApJ},
  \hypersetup{urlcolor=blue}\href{https://ui.adsabs.harvard.edu/abs/2020ApJ...897..114B}{897,
  114}

\bibitem[{{Brooks} \& {Gelman}(1998)}]{Brooks_Gelman_1998}
{Brooks}, S.~P., \& {Gelman}, A. 1998,
  \hypersetup{urlcolor=magenta}\href{https://dx.doi.org/10.1080/10618600.1998.10474787}{Journal
  of Computational and Graphical Statistics}, 7, 434

\bibitem[{{Chambers} {et~al.}(2016){Chambers}, {Magnier}, {Metcalfe},
  {Flewelling}, {Huber}, {Waters}, {Denneau}, {Draper}, {Farrow}, {Finkbeiner},
  {Holmberg}, {Koppenhoefer}, {Price}, {Rest}, {Saglia}, {Schlafly}, {Smartt},
  {Sweeney}, {Wainscoat}, {Burgett}, {Chastel}, {Grav}, {Heasley}, {Hodapp},
  {Jedicke}, {Kaiser}, {Kudritzki}, {Luppino}, {Lupton}, {Monet}, {Morgan},
  {Onaka}, {Shiao}, {Stubbs}, {Tonry}, {White}, {Ba{\~n}ados}, {Bell},
  {Bender}, {Bernard}, {Boegner}, {Boffi}, {Botticella}, {Calamida},
  {Casertano}, {Chen}, {Chen}, {Cole}, {Deacon}, {Frenk}, {Fitzsimmons},
  {Gezari}, {Gibbs}, {Goessl}, {Goggia}, {Gourgue}, {Goldman}, {Grant},
  {Grebel}, {Hambly}, {Hasinger}, {Heavens}, {Heckman}, {Henderson}, {Henning},
  {Holman}, {Hopp}, {Ip}, {Isani}, {Jackson}, {Keyes}, {Koekemoer}, {Kotak},
  {Le}, {Liska}, {Long}, {Lucey}, {Liu}, {Martin}, {Masci}, {McLean}, {Mindel},
  {Misra}, {Morganson}, {Murphy}, {Obaika}, {Narayan}, {Nieto-Santisteban},
  {Norberg}, {Peacock}, {Pier}, {Postman}, {Primak}, {Rae}, {Rai}, {Riess},
  {Riffeser}, {Rix}, {R{\"o}ser}, {Russel}, {Rutz}, {Schilbach}, {Schultz},
  {Scolnic}, {Strolger}, {Szalay}, {Seitz}, {Small}, {Smith}, {Soderblom},
  {Taylor}, {Thomson}, {Taylor}, {Thakar}, {Thiel}, {Thilker}, {Unger},
  {Urata}, {Valenti}, {Wagner}, {Walder}, {Walter}, {Watters}, {Werner},
  {Wood-Vasey}, \& {Wyse}}]{Chambers_2016}
{Chambers}, K.~C., {Magnier}, E.~A., {Metcalfe}, N., {et~al.} 2016, arXiv
  e-prints,
  \hypersetup{urlcolor=magenta}\href{https://arxiv.org/abs/1612.05560}{arXiv}{:}\hypersetup{urlcolor=blue}\href{https://ui.adsabs.harvard.edu/abs/2016arXiv161205560C}{1612.05560}

\bibitem[{{Chen} {et~al.}(2022){Chen}, {Yan}, {Kangas}, {Lunnan}, {Sollerman},
  {Schulze}, {Perley}, {Chen}, {Gal-Yam}, {Wang}, {De}, {Taggart}, {Bellm},
  {Bloom}, {Dekany}, {Graham}, {Kasliwal}, {Kulkarni}, {Laher}, {Neill}, \&
  {Rusholme}}]{Chen_2022b}
{Chen}, Z.~H., {Yan}, L., {Kangas}, T., {et~al.} 2022, arXiv e-prints,
  \hypersetup{urlcolor=magenta}\href{https://arxiv.org/abs/2202.02060}{arXiv}{:}\hypersetup{urlcolor=blue}\href{https://ui.adsabs.harvard.edu/abs/2022arXiv220202060C}{2202.02060}

\bibitem[{Chomiuk {et~al.}(2011)Chomiuk, Chornock, Soderberg, Berger,
  Chevalier, Foley, Huber, Narayan, Rest, Gezari, Kirshner, Riess, Rodney,
  Smartt, Stubbs, Tonry, Wood-Vasey, Burgett, Chambers, Czekala, Flewelling,
  Forster, Kaiser, Kudritzki, Magnier, Martin, Morgan, Neill, Price, Roth,
  Sanders, \& Wainscoat}]{Chomiuk_2011}
Chomiuk, L., Chornock, R., Soderberg, A.~M., {et~al.} 2011,
  \hypersetup{urlcolor=magenta}\href{https://dx.doi.org/10.1088/0004-637x/743/2/114}{ApJ},
  \hypersetup{urlcolor=blue}\href{https://ui.adsabs.harvard.edu/abs/2011ApJ...743..114L}{743,
  114}

\bibitem[{{Curran}(2014)}]{Curran_2014}
{Curran}, P.~A. 2014,
  \hypersetup{urlcolor=magenta}\href{https://arxiv.org/abs/1411.3816}{arXiv}{:}\hypersetup{urlcolor=blue}\href{https://ui.adsabs.harvard.edu/abs/2014arXiv1411.3816C}{1411.3816}

\bibitem[{{De Cia} {et~al.}(2018){De Cia}, {Gal-Yam}, {Rubin}, {Leloudas},
  {Vreeswijk}, {Perley}, {Quimby}, {Yan}, {Sullivan}, {Fl{\"o}rs}, {Sollerman},
  {Bersier}, {Cenko}, {Gal-Yam}, {Maguire}, {Ofek}, {Prentice}, {Schulze},
  {Spyromilio}, {Valenti}, {Arcavi}, {Corsi}, {Howell}, {Mazzali}, {Kasliwal},
  {Taddia}, \& {Yaron}}]{DeCia_2018}
{De Cia}, A., {Gal-Yam}, A., {Rubin}, A., {et~al.} 2018,
  \hypersetup{urlcolor=magenta}\href{https://dx.doi.org/10.3847/1538-4357/aab9b6}{ApJ},
  \hypersetup{urlcolor=blue}\href{https://ui.adsabs.harvard.edu/abs/2018ApJ...860..100D}{860,
  100}

\bibitem[{Dessart {et~al.}(2012)Dessart, Hillier, Waldman, Livne, \&
  Blondin}]{Dessart_2012}
Dessart, L., Hillier, D.~J., Waldman, R., Livne, E., \& Blondin, S. 2012,
  \hypersetup{urlcolor=magenta}\href{https://dx.doi.org/10.1111/j.1745-3933.2012.01329.x}{MNRAS},
  \hypersetup{urlcolor=blue}\href{https://ui.adsabs.harvard.edu/abs/2012MNRAS.426L..76D}{426,
  L76}

\bibitem[{{Fitzpatrick}(1999)}]{Fitzpatrick_1999}
{Fitzpatrick}, E.~L. 1999,
  \hypersetup{urlcolor=magenta}\href{https://dx.doi.org/10.1086/316293}{\pasp},
  \hypersetup{urlcolor=blue}\href{https://ui.adsabs.harvard.edu/abs/1999PASP..111...63F}{111,
  63}

\bibitem[{{Frohmaier} {et~al.}(2021){Frohmaier}, {Angus}, {Vincenzi},
  {Sullivan}, {Smith}, {Nugent}, {Cenko}, {Gal-Yam}, {Kulkarni}, {Law}, \&
  {Quimby}}]{Frohmaier_2021}
{Frohmaier}, C., {Angus}, C.~R., {Vincenzi}, M., {et~al.} 2021,
  \hypersetup{urlcolor=magenta}\href{https://dx.doi.org/10.1093/mnras/staa3607}{\mnras},
  \hypersetup{urlcolor=blue}\href{https://ui.adsabs.harvard.edu/abs/2021MNRAS.500.5142F}{500,
  5142}

\bibitem[{Gelman \& Rubin(1992)}]{Gelman_1992}
Gelman, A., \& Rubin, D.~B. 1992,
  \hypersetup{urlcolor=magenta}\href{https://dx.doi.org/10.1214/ss/1177011136}{StaSc},
  \hypersetup{urlcolor=blue}\href{https://ui.adsabs.harvard.edu/abs/1992StaSc...7..457G}{7,
  457}

\bibitem[{{Gomez} {et~al.}(2021){Gomez}, {Berger}, {Hosseinzadeh}, {Blanchard},
  {Nicholl}, \& {Villar}}]{Gomez_2021}
{Gomez}, S., {Berger}, E., {Hosseinzadeh}, G., {et~al.} 2021,
  \hypersetup{urlcolor=magenta}\href{https://dx.doi.org/10.3847/1538-4357/abf5e3}{\apj},
  \hypersetup{urlcolor=blue}\href{https://ui.adsabs.harvard.edu/abs/2021ApJ...913..143G}{913,
  143}

\bibitem[{Guillochon {et~al.}(2018)Guillochon, Nicholl, Villar, Mockler,
  Narayan, Mandel, Berger, \& Williams}]{Guillochon_2018}
Guillochon, J., Nicholl, M., Villar, V.~A., {et~al.} 2018,
  \hypersetup{urlcolor=magenta}\href{https://dx.doi.org/10.3847/1538-4365/aab761}{ApJS},
  \hypersetup{urlcolor=blue}\href{https://ui.adsabs.harvard.edu/abs/2018ApJS..236....6G}{236,
  6}

\bibitem[{{Hosseinzadeh} {et~al.}(2021){Hosseinzadeh}, {Berger}, {Metzger},
  {Gomez}, {Nicholl}, \& {Blanchard}}]{Hosseinzadeh_2021}
{Hosseinzadeh}, G., {Berger}, E., {Metzger}, B.~D., {et~al.} 2021, arXiv
  e-prints,
  \hypersetup{urlcolor=magenta}\href{https://arxiv.org/abs/2109.09743}{arXiv}{:}\hypersetup{urlcolor=blue}\href{https://ui.adsabs.harvard.edu/abs/2021arXiv210909743H}{2109.09743}

\bibitem[{{Hosseinzadeh} {et~al.}(2020){Hosseinzadeh}, {Dauphin}, {Villar},
  {Berger}, {Jones}, {Challis}, {Chornock}, {Drout}, {Foley}, {Kirshner},
  {Lunnan}, {Margutti}, {Milisavljevic}, {Pan}, {Rest}, {Scolnic}, {Magnier},
  {Metcalfe}, {Wainscoat}, \& {Waters}}]{Hosseinzadeh_2020}
{Hosseinzadeh}, G., {Dauphin}, F., {Villar}, V.~A., {et~al.} 2020,
  \hypersetup{urlcolor=magenta}\href{https://dx.doi.org/10.3847/1538-4357/abc42b}{\apj},
  \hypersetup{urlcolor=blue}\href{https://ui.adsabs.harvard.edu/abs/2020ApJ...905...93H}{905,
  93}

\bibitem[{{Hsu} {et~al.}(2021){Hsu}, {Hosseinzadeh}, \& {Berger}}]{Hsu_2021}
{Hsu}, B., {Hosseinzadeh}, G., \& {Berger}, E. 2021,
  \hypersetup{urlcolor=magenta}\href{https://dx.doi.org/10.3847/1538-4357/ac1aca}{\apj},
  \hypersetup{urlcolor=blue}\href{https://ui.adsabs.harvard.edu/abs/2021ApJ...921..180H}{921,
  180}

\bibitem[{{Huber} {et~al.}(2017){Huber}, {PS1 Science Consortium}, \&
  {Pan-STARRS IPP Team}}]{Huber_2017}
{Huber}, M., {PS1 Science Consortium}, \& {Pan-STARRS IPP Team}. 2017, in
  American Astronomical Society Meeting Abstracts, Vol. 229, American
  Astronomical Society Meeting Abstracts \#229, 237.06

\bibitem[{Hunter(2007)}]{Hunter_2007}
Hunter, J.~D. 2007,
  \hypersetup{urlcolor=magenta}\href{https://dx.doi.org/10.1109/MCSE.2007.55}{CSE},
  \hypersetup{urlcolor=blue}\href{https://ui.adsabs.harvard.edu/abs/2007CSE.....9...90H}{9,
  90}

\bibitem[{{Inserra} {et~al.}(2013){Inserra}, {Smartt}, {Jerkstrand}, {Valenti},
  {Fraser}, {Wright}, {Smith}, {Chen}, {Kotak}, {Pastorello}, {Nicholl},
  {Bresolin}, {Kudritzki}, {Benetti}, {Botticella}, {Burgett}, {Chambers},
  {Ergon}, {Flewelling}, {Fynbo}, {Geier}, {Hodapp}, {Howell}, {Huber},
  {Kaiser}, {Leloudas}, {Magill}, {Magnier}, {McCrum}, {Metcalfe}, {Price},
  {Rest}, {Sollerman}, {Sweeney}, {Taddia}, {Taubenberger}, {Tonry},
  {Wainscoat}, {Waters}, \& {Young}}]{Inserra_2013}
{Inserra}, C., {Smartt}, S.~J., {Jerkstrand}, A., {et~al.} 2013,
  \hypersetup{urlcolor=magenta}\href{https://dx.doi.org/10.1088/0004-637X/770/2/128}{\apj},
  \hypersetup{urlcolor=blue}\href{https://ui.adsabs.harvard.edu/abs/2013ApJ...770..128I}{770,
  128}

\bibitem[{{Inserra} {et~al.}(2017){Inserra}, {Nicholl}, {Chen}, {Jerkstrand},
  {Smartt}, {Kr{\"u}hler}, {Anderson}, {Baltay}, {Della Valle}, {Fraser},
  {Gal-Yam}, {Galbany}, {Kankare}, {Maguire}, {Rabinowitz}, {Smith}, {Valenti},
  \& {Young}}]{Inserra_2017}
{Inserra}, C., {Nicholl}, M., {Chen}, T.~W., {et~al.} 2017,
  \hypersetup{urlcolor=magenta}\href{https://dx.doi.org/10.1093/mnras/stx834}{MNRAS},
  \hypersetup{urlcolor=blue}\href{https://ui.adsabs.harvard.edu/abs/2017MNRAS.468.4642I}{468,
  4642}

\bibitem[{{Ivezi{\'c}} {et~al.}(2019){Ivezi{\'c}}, {Kahn}, {Tyson}, {Abel},
  {Acosta}, {Allsman}, {Alonso}, {AlSayyad}, {Anderson}, {Andrew}, {Angel},
  {Angeli}, {Ansari}, {Antilogus}, {Araujo}, {Armstrong}, {Arndt}, {Astier},
  {Aubourg}, {Auza}, {Axelrod}, {Bard}, {Barr}, {Barrau}, {Bartlett}, {Bauer},
  {Bauman}, {Baumont}, {Bechtol}, {Bechtol}, {Becker}, {Becla}, {Beldica},
  {Bellavia}, {Bianco}, {Biswas}, {Blanc}, {Blazek}, {Blandford}, {Bloom},
  {Bogart}, {Bond}, {Booth}, {Borgland}, {Borne}, {Bosch}, {Boutigny},
  {Brackett}, {Bradshaw}, {Brandt}, {Brown}, {Bullock}, {Burchat}, {Burke},
  {Cagnoli}, {Calabrese}, {Callahan}, {Callen}, {Carlin}, {Carlson},
  {Chandrasekharan}, {Charles-Emerson}, {Chesley}, {Cheu}, {Chiang}, {Chiang},
  {Chirino}, {Chow}, {Ciardi}, {Claver}, {Cohen-Tanugi}, {Cockrum}, {Coles},
  {Connolly}, {Cook}, {Cooray}, {Covey}, {Cribbs}, {Cui}, {Cutri}, {Daly},
  {Daniel}, {Daruich}, {Daubard}, {Daues}, {Dawson}, {Delgado}, {Dellapenna},
  {de Peyster}, {de Val-Borro}, {Digel}, {Doherty}, {Dubois},
  {Dubois-Felsmann}, {Durech}, {Economou}, {Eifler}, {Eracleous}, {Emmons},
  {Fausti Neto}, {Ferguson}, {Figueroa}, {Fisher-Levine}, {Focke}, {Foss},
  {Frank}, {Freemon}, {Gangler}, {Gawiser}, {Geary}, {Gee}, {Geha}, {Gessner},
  {Gibson}, {Gilmore}, {Glanzman}, {Glick}, {Goldina}, {Goldstein}, {Goodenow},
  {Graham}, {Gressler}, {Gris}, {Guy}, {Guyonnet}, {Haller}, {Harris},
  {Hascall}, {Haupt}, {Hernandez}, {Herrmann}, {Hileman}, {Hoblitt}, {Hodgson},
  {Hogan}, {Howard}, {Huang}, {Huffer}, {Ingraham}, {Innes}, {Jacoby}, {Jain},
  {Jammes}, {Jee}, {Jenness}, {Jernigan}, {Jevremovi{\'c}}, {Johns}, {Johnson},
  {Johnson}, {Jones}, {Juramy-Gilles}, {Juri{\'c}}, {Kalirai}, {Kallivayalil},
  {Kalmbach}, {Kantor}, {Karst}, {Kasliwal}, {Kelly}, {Kessler}, {Kinnison},
  {Kirkby}, {Knox}, {Kotov}, {Krabbendam}, {Krughoff}, {Kub{\'a}nek},
  {Kuczewski}, {Kulkarni}, {Ku}, {Kurita}, {Lage}, {Lambert}, {Lange},
  {Langton}, {Le Guillou}, {Levine}, {Liang}, {Lim}, {Lintott}, {Long},
  {Lopez}, {Lotz}, {Lupton}, {Lust}, {MacArthur}, {Mahabal}, {Mandelbaum},
  {Markiewicz}, {Marsh}, {Marshall}, {Marshall}, {May}, {McKercher}, {McQueen},
  {Meyers}, {Migliore}, {Miller}, {Mills}, {Miraval}, {Moeyens}, {Moolekamp},
  {Monet}, {Moniez}, {Monkewitz}, {Montgomery}, {Morrison}, {Mueller},
  {Muller}, {Mu{\~n}oz Arancibia}, {Neill}, {Newbry}, {Nief}, {Nomerotski},
  {Nordby}, {O'Connor}, {Oliver}, {Olivier}, {Olsen}, {O'Mullane}, {Ortiz},
  {Osier}, {Owen}, {Pain}, {Palecek}, {Parejko}, {Parsons}, {Pease},
  {Peterson}, {Peterson}, {Petravick}, {Libby Petrick}, {Petry},
  {Pierfederici}, {Pietrowicz}, {Pike}, {Pinto}, {Plante}, {Plate}, {Plutchak},
  {Price}, {Prouza}, {Radeka}, {Rajagopal}, {Rasmussen}, {Regnault}, {Reil},
  {Reiss}, {Reuter}, {Ridgway}, {Riot}, {Ritz}, {Robinson}, {Roby}, {Roodman},
  {Rosing}, {Roucelle}, {Rumore}, {Russo}, {Saha}, {Sassolas}, {Schalk},
  {Schellart}, {Schindler}, {Schmidt}, {Schneider}, {Schneider}, {Schoening},
  {Schumacher}, {Schwamb}, {Sebag}, {Selvy}, {Sembroski}, {Seppala}, {Serio},
  {Serrano}, {Shaw}, {Shipsey}, {Sick}, {Silvestri}, {Slater}, {Smith},
  {Smith}, {Sobhani}, {Soldahl}, {Storrie-Lombardi}, {Stover}, {Strauss},
  {Street}, {Stubbs}, {Sullivan}, {Sweeney}, {Swinbank}, {Szalay}, {Takacs},
  {Tether}, {Thaler}, {Thayer}, {Thomas}, {Thornton}, {Thukral}, {Tice},
  {Trilling}, {Turri}, {Van Berg}, {Vanden Berk}, {Vetter}, {Virieux},
  {Vucina}, {Wahl}, {Walkowicz}, {Walsh}, {Walter}, {Wang}, {Wang}, {Warner},
  {Wiecha}, {Willman}, {Winters}, {Wittman}, {Wolff}, {Wood-Vasey}, {Wu},
  {Xin}, {Yoachim}, \& {Zhan}}]{Ivezic_2019}
{Ivezi{\'c}}, {\v{Z}}., {Kahn}, S.~M., {Tyson}, J.~A., {et~al.} 2019,
  \hypersetup{urlcolor=magenta}\href{https://dx.doi.org/10.3847/1538-4357/ab042c}{\apj},
  \hypersetup{urlcolor=blue}\href{https://ui.adsabs.harvard.edu/abs/2019ApJ...873..111I}{873,
  111}

\bibitem[{{Jerkstrand} {et~al.}(2017){Jerkstrand}, {Smartt}, {Inserra},
  {Nicholl}, {Chen}, {Kr{\"u}hler}, {Sollerman}, {Taubenberger}, {Gal-Yam},
  {Kankare}, {Maguire}, {Fraser}, {Valenti}, {Sullivan}, {Cartier}, \&
  {Young}}]{Jerkstrand_2017}
{Jerkstrand}, A., {Smartt}, S.~J., {Inserra}, C., {et~al.} 2017,
  \hypersetup{urlcolor=magenta}\href{https://dx.doi.org/10.3847/1538-4357/835/1/13}{ApJ},
  \hypersetup{urlcolor=blue}\href{https://ui.adsabs.harvard.edu/abs/2017ApJ...835...13J}{835,
  13}

\bibitem[{{Kasen} \& {Bildsten}(2010)}]{Kasen_Bildsten_2010}
{Kasen}, D., \& {Bildsten}, L. 2010,
  \hypersetup{urlcolor=magenta}\href{https://dx.doi.org/10.1088/0004-637X/717/1/245}{ApJ},
  \hypersetup{urlcolor=blue}\href{https://ui.adsabs.harvard.edu/abs/2010ApJ...717..245K}{717,
  245}

\bibitem[{{Lunnan} {et~al.}(2013){Lunnan}, {Chornock}, {Berger},
  {Milisavljevic}, {Drout}, {Sanders}, {Challis}, {Czekala}, {Foley}, {Fong},
  {Huber}, {Kirshner}, {Leibler}, {Marion}, {McCrum}, {Narayan}, {Rest},
  {Roth}, {Scolnic}, {Smartt}, {Smith}, {Soderberg}, {Stubbs}, {Tonry},
  {Burgett}, {Chambers}, {Kudritzki}, {Magnier}, \& {Price}}]{Lunnan_2013}
{Lunnan}, R., {Chornock}, R., {Berger}, E., {et~al.} 2013,
  \hypersetup{urlcolor=magenta}\href{https://dx.doi.org/10.1088/0004-637X/771/2/97}{ApJ},
  \hypersetup{urlcolor=blue}\href{https://ui.adsabs.harvard.edu/abs/2013ApJ...771...97L}{771,
  97}

\bibitem[{{Lunnan} {et~al.}(2014){Lunnan}, {Chornock}, {Berger}, {Laskar},
  {Fong}, {Rest}, {Sanders}, {Challis}, {Drout}, {Foley}, {Huber}, {Kirshner},
  {Leibler}, {Marion}, {McCrum}, {Milisavljevic}, {Narayan}, {Scolnic},
  {Smartt}, {Smith}, {Soderberg}, {Tonry}, {Burgett}, {Chambers}, {Flewelling},
  {Hodapp}, {Kaiser}, {Magnier}, {Price}, \& {Wainscoat}}]{Lunnan_2014}
{Lunnan}, R., {Chornock}, R., {Berger}, E., {et~al.} 2014,
  \hypersetup{urlcolor=magenta}\href{https://dx.doi.org/10.1088/0004-637X/787/2/138}{ApJ},
  \hypersetup{urlcolor=blue}\href{https://ui.adsabs.harvard.edu/abs/2014ApJ...787..138L}{787,
  138}

\bibitem[{Lunnan {et~al.}(2018)Lunnan, Chornock, Berger, Jones, Rest, Czekala,
  Dittmann, Drout, Foley, Fong, Kirshner, Laskar, Leibler, Margutti,
  Milisavljevic, Narayan, Pan, Riess, Roth, Sanders, Scolnic, Smartt, Smith,
  Chambers, Draper, Flewelling, Huber, Kaiser, Kudritzki, Magnier, Metcalfe,
  Wainscoat, Waters, \& Willman}]{Lunnan_2018}
Lunnan, R., Chornock, R., Berger, E., {et~al.} 2018,
  \hypersetup{urlcolor=magenta}\href{https://dx.doi.org/10.3847/1538-4357/aa9f1a}{ApJ},
  \hypersetup{urlcolor=blue}\href{https://ui.adsabs.harvard.edu/abs/2018ApJ...852...81R}{852,
  81}

\bibitem[{{Margalit} {et~al.}(2018){Margalit}, {Metzger}, {Thompson},
  {Nicholl}, \& {Sukhbold}}]{Margalit_2018}
{Margalit}, B., {Metzger}, B.~D., {Thompson}, T.~A., {Nicholl}, M., \&
  {Sukhbold}, T. 2018,
  \hypersetup{urlcolor=magenta}\href{https://dx.doi.org/10.1093/mnras/sty013}{\mnras},
  \hypersetup{urlcolor=blue}\href{https://ui.adsabs.harvard.edu/abs/2018MNRAS.475.2659M}{475,
  2659}

\bibitem[{{Mazzali} {et~al.}(2016){Mazzali}, {Sullivan}, {Pian}, {Greiner}, \&
  {Kann}}]{Mazzali_2016}
{Mazzali}, P.~A., {Sullivan}, M., {Pian}, E., {Greiner}, J., \& {Kann}, D.~A.
  2016,
  \hypersetup{urlcolor=magenta}\href{https://dx.doi.org/10.1093/mnras/stw512}{MNRAS},
  \hypersetup{urlcolor=blue}\href{https://ui.adsabs.harvard.edu/abs/2016MNRAS.458.3455M}{458,
  3455}

\bibitem[{{Metzger} {et~al.}(2015){Metzger}, {Margalit}, {Kasen}, \&
  {Quataert}}]{Metzger_2015}
{Metzger}, B.~D., {Margalit}, B., {Kasen}, D., \& {Quataert}, E. 2015,
  \hypersetup{urlcolor=magenta}\href{https://dx.doi.org/10.1093/mnras/stv2224}{MNRAS},
  \hypersetup{urlcolor=blue}\href{https://ui.adsabs.harvard.edu/abs/2015MNRAS.454.3311M}{454,
  3311}

\bibitem[{{Nicholl}(2021)}]{Nicholl_2021}
{Nicholl}, M. 2021,
  \hypersetup{urlcolor=magenta}\href{https://dx.doi.org/10.1093/astrogeo/atab092}{Astronomy
  and Geophysics},
  \hypersetup{urlcolor=blue}\href{https://ui.adsabs.harvard.edu/abs/2021A&G....62.5.34N}{62,
  5.34}

\bibitem[{{Nicholl} {et~al.}(2019){Nicholl}, {Berger}, {Blanchard}, {Gomez}, \&
  {Chornock}}]{Nicholl_2019}
{Nicholl}, M., {Berger}, E., {Blanchard}, P.~K., {Gomez}, S., \& {Chornock}, R.
  2019,
  \hypersetup{urlcolor=magenta}\href{https://dx.doi.org/10.3847/1538-4357/aaf470}{ApJ},
  \hypersetup{urlcolor=blue}\href{https://ui.adsabs.harvard.edu/abs/2019ApJ...871..102N}{871,
  102}

\bibitem[{{Nicholl} {et~al.}(2017{\natexlab{\hspace{0pt}a}}){Nicholl},
  {Berger}, {Margutti}, {Blanchard}, {Milisavljevic}, {Challis}, {Metzger}, \&
  {Chornock}}]{Nicholl_2017c}
{Nicholl}, M., {Berger}, E., {Margutti}, R., {et~al.}
  2017{\natexlab{\hspace{0pt}a}},
  \hypersetup{urlcolor=magenta}\href{https://dx.doi.org/10.3847/2041-8213/aa56c5}{ApJL},
  \hypersetup{urlcolor=blue}\href{https://ui.adsabs.harvard.edu/abs/2017ApJ...835L...8N}{835,
  L8}

\bibitem[{{Nicholl} {et~al.}(2017{\natexlab{\hspace{0pt}b}}){Nicholl},
  {Guillochon}, \& {Berger}}]{Nicholl_2017b}
{Nicholl}, M., {Guillochon}, J., \& {Berger}, E.
  2017{\natexlab{\hspace{0pt}b}},
  \hypersetup{urlcolor=magenta}\href{https://dx.doi.org/10.3847/1538-4357/aa9334}{\apj},
  \hypersetup{urlcolor=blue}\href{https://ui.adsabs.harvard.edu/abs/2017ApJ...850...55N}{850,
  55}

\bibitem[{{Nicholl} {et~al.}(2014){Nicholl}, {Smartt}, {Jerkstrand}, {Inserra},
  {Anderson}, {Baltay}, {Benetti}, {Chen}, {Elias-Rosa}, {Feindt}, {Fraser},
  {Gal-Yam}, {Hadjiyska}, {Howell}, {Kotak}, {Lawrence}, {Leloudas},
  {Margheim}, {Mattila}, {McCrum}, {McKinnon}, {Mead}, {Nugent}, {Rabinowitz},
  {Rest}, {Smith}, {Sollerman}, {Sullivan}, {Taddia}, {Valenti}, {Walker}, \&
  {Young}}]{Nicholl_2014}
{Nicholl}, M., {Smartt}, S.~J., {Jerkstrand}, A., {et~al.} 2014,
  \hypersetup{urlcolor=magenta}\href{https://dx.doi.org/10.1093/mnras/stu1579}{\mnras},
  \hypersetup{urlcolor=blue}\href{https://ui.adsabs.harvard.edu/abs/2014MNRAS.444.2096N}{444,
  2096}

\bibitem[{{Nicholl} {et~al.}(2015){Nicholl}, {Smartt}, {Jerkstrand}, {Inserra},
  {Sim}, {Chen}, {Benetti}, {Fraser}, {Gal-Yam}, {Kankare}, {Maguire}, {Smith},
  {Sullivan}, {Valenti}, {Young}, {Baltay}, {Bauer}, {Baumont}, {Bersier},
  {Botticella}, {Childress}, {Dennefeld}, {Della Valle}, {Elias-Rosa},
  {Feindt}, {Galbany}, {Hadjiyska}, {Le Guillou}, {Leloudas}, {Mazzali},
  {McKinnon}, {Polshaw}, {Rabinowitz}, {Rostami}, {Scalzo}, {Schmidt},
  {Schulze}, {Sollerman}, {Taddia}, \& {Yuan}}]{Nicholl_2015b}
{Nicholl}, M., {Smartt}, S.~J., {Jerkstrand}, A., {et~al.} 2015,
  \hypersetup{urlcolor=magenta}\href{https://dx.doi.org/10.1093/mnras/stv1522}{MNRAS},
  \hypersetup{urlcolor=blue}\href{https://ui.adsabs.harvard.edu/abs/2015MNRAS.452.3869N}{452,
  3869}

\bibitem[{{Nicholl} {et~al.}(2016{\natexlab{\hspace{0pt}a}}){Nicholl},
  {Berger}, {Smartt}, {Margutti}, {Kamble}, {Alexander}, {Chen}, {Inserra},
  {Arcavi}, {Blanchard}, {Cartier}, {Chambers}, {Childress}, {Chornock},
  {Cowperthwaite}, {Drout}, {Flewelling}, {Fraser}, {Gal-Yam}, {Galbany},
  {Harmanen}, {Holoien}, {Hosseinzadeh}, {Howell}, {Huber}, {Jerkstrand },
  {Kankare}, {Kochanek}, {Lin}, {Lunnan}, {Magnier}, {Maguire}, {McCully},
  {McDonald}, {Metzger}, {Milisavljevic}, {Mitra}, {Reynolds}, {Saario},
  {Shappee}, {Smith}, {Valenti}, {Villar}, {Waters}, \&
  {Young}}]{Nicholl_2016a}
{Nicholl}, M., {Berger}, E., {Smartt}, S.~J., {et~al.}
  2016{\natexlab{\hspace{0pt}a}},
  \hypersetup{urlcolor=magenta}\href{https://dx.doi.org/10.3847/0004-637X/826/1/39}{ApJ},
  \hypersetup{urlcolor=blue}\href{https://ui.adsabs.harvard.edu/abs/2016ApJ...826...39N}{826,
  39}

\bibitem[{{Nicholl} {et~al.}(2016{\natexlab{\hspace{0pt}b}}){Nicholl},
  {Berger}, {Margutti}, {Chornock}, {Blanchard}, {Jerkstrand}, {Smartt},
  {Arcavi}, {Challis}, {Chambers}, {Chen}, {Cowperthwaite}, {Gal-Yam},
  {Hosseinzadeh}, {Howell}, {Inserra}, {Kankare}, {Magnier}, {Maguire},
  {Mazzali}, {McCully}, {Milisavljevic}, {Smith}, {Taubenberger}, {Valenti},
  {Wainscoat}, {Yaron}, \& {Young}}]{Nicholl_2016b}
{Nicholl}, M., {Berger}, E., {Margutti}, R., {et~al.}
  2016{\natexlab{\hspace{0pt}b}},
  \hypersetup{urlcolor=magenta}\href{https://dx.doi.org/10.3847/2041-8205/828/2/L18}{ApJL},
  \hypersetup{urlcolor=blue}\href{https://ui.adsabs.harvard.edu/abs/2016ApJ...828L..18N}{828,
  L18}

\bibitem[{{Nicholl} {et~al.}(2018){Nicholl}, {Blanchard}, {Berger},
  {Alexander}, {Metzger}, {Bhirombhakdi}, {Chornock}, {Coppejans}, {Gomez},
  {Margalit}, {Margutti}, \& {Terreran}}]{Nicholl_2018}
{Nicholl}, M., {Blanchard}, P.~K., {Berger}, E., {et~al.} 2018,
  \hypersetup{urlcolor=magenta}\href{https://dx.doi.org/10.3847/2041-8213/aae70d}{ApJL},
  \hypersetup{urlcolor=blue}\href{https://ui.adsabs.harvard.edu/abs/2018ApJ...866L..24N}{866,
  L24}

\bibitem[{Oliphant(2006)}]{Oliphant_2006}
Oliphant, T.~E. 2006, A guide to {NumPy} (USA: Trelgol Publishing)

\bibitem[{{Perley} {et~al.}(2016){Perley}, {Quimby}, {Yan}, {Vreeswijk}, {De
  Cia}, {Lunnan}, {Gal-Yam}, {Yaron}, {Filippenko}, {Graham}, {Laher}, \&
  {Nugent}}]{Perley_2016}
{Perley}, D.~A., {Quimby}, R.~M., {Yan}, L., {et~al.} 2016,
  \hypersetup{urlcolor=magenta}\href{https://dx.doi.org/10.3847/0004-637X/830/1/13}{ApJ},
  \hypersetup{urlcolor=blue}\href{https://ui.adsabs.harvard.edu/abs/2016ApJ...830...13P}{830,
  13}

\bibitem[{{Perley} {et~al.}(2020){Perley}, {Fremling}, {Sollerman}, {Miller},
  {Dahiwale}, {Sharma}, {Bellm}, {Biswas}, {Brink}, {Bruch}, {De}, {Dekany},
  {Drake}, {Duev}, {Filippenko}, {Gal-Yam}, {Goobar}, {Graham}, {Graham}, {Ho},
  {Irani}, {Kasliwal}, {Kim}, {Kulkarni}, {Mahabal}, {Masci}, {Modak}, {Neill},
  {Nordin}, {Riddle}, {Soumagnac}, {Strotjohann}, {Schulze}, {Taggart},
  {Tzanidakis}, {Walters}, \& {Yan}}]{Perley_2020}
{Perley}, D.~A., {Fremling}, C., {Sollerman}, J., {et~al.} 2020,
  \hypersetup{urlcolor=magenta}\href{https://dx.doi.org/10.3847/1538-4357/abbd98}{\apj},
  \hypersetup{urlcolor=blue}\href{https://ui.adsabs.harvard.edu/abs/2020ApJ...904...35P}{904,
  35}

\bibitem[{{Planck Collaboration} {et~al.}(2016){Planck Collaboration}, {Ade},
  {Aghanim}, {Arnaud}, {Ashdown}, {Aumont}, {Baccigalupi}, {Banday},
  {Barreiro}, {Bartlett}, {Bartolo}, {Battaner}, {Battye}, {Benabed},
  {Beno{\^\i}t}, {Benoit-L{\'e}vy}, {Bernard}, {Bersanelli}, {Bielewicz},
  {Bock}, {Bonaldi}, {Bonavera}, {Bond}, {Borrill}, {Bouchet}, {Boulanger},
  {Bucher}, {Burigana}, {Butler}, {Calabrese}, {Cardoso}, {Catalano},
  {Challinor}, {Chamballu}, {Chary}, {Chiang}, {Chluba}, {Christensen},
  {Church}, {Clements}, {Colombi}, {Colombo}, {Combet}, {Coulais}, {Crill},
  {Curto}, {Cuttaia}, {Danese}, {Davies}, {Davis}, {de Bernardis}, {de Rosa},
  {de Zotti}, {Delabrouille}, {D{\'e}sert}, {Di Valentino}, {Dickinson},
  {Diego}, {Dolag}, {Dole}, {Donzelli}, {Dor{\'e}}, {Douspis}, {Ducout},
  {Dunkley}, {Dupac}, {Efstathiou}, {Elsner}, {En{\ss}lin}, {Eriksen},
  {Farhang}, {Fergusson}, {Finelli}, {Forni}, {Frailis}, {Fraisse},
  {Franceschi}, {Frejsel}, {Galeotta}, {Galli}, {Ganga}, {Gauthier}, {Gerbino},
  {Ghosh}, {Giard}, {Giraud-H{\'e}raud}, {Giusarma}, {Gjerl{\o}w},
  {Gonz{\'a}lez-Nuevo}, {G{\'o}rski}, {Gratton}, {Gregorio}, {Gruppuso},
  {Gudmundsson}, {Hamann}, {Hansen}, {Hanson}, {Harrison}, {Helou},
  {Henrot-Versill{\'e}}, {Hern{\'a}ndez-Monteagudo}, {Herranz}, {Hildebrandt},
  {Hivon}, {Hobson}, {Holmes}, {Hornstrup}, {Hovest}, {Huang}, {Huffenberger},
  {Hurier}, {Jaffe}, {Jaffe}, {Jones}, {Juvela}, {Keih{\"a}nen}, {Keskitalo},
  {Kisner}, {Kneissl}, {Knoche}, {Knox}, {Kunz}, {Kurki-Suonio}, {Lagache},
  {L{\"a}hteenm{\"a}ki}, {Lamarre}, {Lasenby}, {Lattanzi}, {Lawrence}, {Leahy},
  {Leonardi}, {Lesgourgues}, {Levrier}, {Lewis}, {Liguori}, {Lilje},
  {Linden-V{\o}rnle}, {L{\'o}pez-Caniego}, {Lubin}, {Mac{\'\i}as-P{\'e}rez},
  {Maggio}, {Maino}, {Mandolesi}, {Mangilli}, {Marchini}, {Maris}, {Martin},
  {Martinelli}, {Mart{\'\i}nez-Gonz{\'a}lez}, {Masi}, {Matarrese}, {McGehee},
  {Meinhold}, {Melchiorri}, {Melin}, {Mendes}, {Mennella}, {Migliaccio},
  {Millea}, {Mitra}, {Miville-Desch{\^e}nes}, {Moneti}, {Montier}, {Morgante},
  {Mortlock}, {Moss}, {Munshi}, {Murphy}, {Naselsky}, {Nati}, {Natoli},
  {Netterfield}, {N{\o}rgaard-Nielsen}, {Noviello}, {Novikov}, {Novikov},
  {Oxborrow}, {Paci}, {Pagano}, {Pajot}, {Paladini}, {Paoletti}, {Partridge},
  {Pasian}, {Patanchon}, {Pearson}, {Perdereau}, {Perotto}, {Perrotta},
  {Pettorino}, {Piacentini}, {Piat}, {Pierpaoli}, {Pietrobon}, {Plaszczynski},
  {Pointecouteau}, {Polenta}, {Popa}, {Pratt}, {Pr{\'e}zeau}, {Prunet},
  {Puget}, {Rachen}, {Reach}, {Rebolo}, {Reinecke}, {Remazeilles}, {Renault},
  {Renzi}, {Ristorcelli}, {Rocha}, {Rosset}, {Rossetti}, {Roudier},
  {Rouill{\'e} d'Orfeuil}, {Rowan-Robinson}, {Rubi{\~n}o-Mart{\'\i}n},
  {Rusholme}, {Said}, {Salvatelli}, {Salvati}, {Sandri}, {Santos},
  {Savelainen}, {Savini}, {Scott}, {Seiffert}, {Serra}, {Shellard}, {Spencer},
  {Spinelli}, {Stolyarov}, {Stompor}, {Sudiwala}, {Sunyaev}, {Sutton},
  {Suur-Uski}, {Sygnet}, {Tauber}, {Terenzi}, {Toffolatti}, {Tomasi},
  {Tristram}, {Trombetti}, {Tucci}, {Tuovinen}, {T{\"u}rler}, {Umana},
  {Valenziano}, {Valiviita}, {Van Tent}, {Vielva}, {Villa}, {Wade}, {Wandelt},
  {Wehus}, {White}, {White}, {Wilkinson}, {Yvon}, {Zacchei}, \&
  {Zonca}}]{Planck_2016}
{Planck Collaboration}, {Ade}, P.~A.~R., {Aghanim}, N., {et~al.} 2016,
  \hypersetup{urlcolor=magenta}\href{https://dx.doi.org/10.1051/0004-6361/201525830}{A\&A},
  \hypersetup{urlcolor=blue}\href{https://ui.adsabs.harvard.edu/abs/2016A&A...594A..13P}{594,
  A13}

\bibitem[{{Privon} {et~al.}(2020){Privon}, {Ricci}, {Aalto}, {Viti}, {Armus},
  {D{\'\i}az-Santos}, {Gonz{\'a}lez-Alfonso}, {Iwasawa}, {Jeff}, {Treister},
  {Bauer}, {Evans}, {Garg}, {Herrero-Illana}, {Mazzarella}, {Larson}, {Blecha},
  {Barcos-Mu{\~n}oz}, {Charmandaris}, {Stierwalt}, \&
  {P{\'e}rez-Torres}}]{Privon_2020}
{Privon}, G.~C., {Ricci}, C., {Aalto}, S., {et~al.} 2020,
  \hypersetup{urlcolor=magenta}\href{https://dx.doi.org/10.3847/1538-4357/ab8015}{\apj},
  \hypersetup{urlcolor=blue}\href{https://ui.adsabs.harvard.edu/abs/2020ApJ...893..149P}{893,
  149}

\bibitem[{{Quimby} {et~al.}(2011){Quimby}, {Kulkarni}, {Kasliwal}, {Gal-Yam},
  {Arcavi}, {Sullivan}, {Nugent}, {Thomas}, {Howell}, {Nakar}, {Bildsten},
  {Theissen}, {Law}, {Dekany}, {Rahmer}, {Hale}, {Smith}, {Ofek}, {Zolkower},
  {Velur}, {Walters}, {Henning}, {Bui}, {McKenna}, {Poznanski}, {Cenko}, \&
  {Levitan}}]{Quimby_2011}
{Quimby}, R.~M., {Kulkarni}, S.~R., {Kasliwal}, M.~M., {et~al.} 2011,
  \hypersetup{urlcolor=magenta}\href{https://dx.doi.org/10.1038/nature10095}{Natur},
  \hypersetup{urlcolor=blue}\href{https://ui.adsabs.harvard.edu/abs/2011Natur.474..487Q}{474,
  487}

\bibitem[{{Quimby} {et~al.}(2018){Quimby}, {De Cia}, {Gal-Yam}, {Leloudas},
  {Lunnan}, {Perley}, {Vreeswijk}, {Yan}, {Bloom}, {Cenko}, {Cooke}, {Ellis},
  {Filippenko}, {Kasliwal}, {Kleiser}, {Kulkarni}, {Matheson}, {Nugent}, {Pan},
  {Silverman}, {Sternberg}, {Sullivan}, \& {Yaron}}]{Quimby_2018}
{Quimby}, R.~M., {De Cia}, A., {Gal-Yam}, A., {et~al.} 2018,
  \hypersetup{urlcolor=magenta}\href{https://dx.doi.org/10.3847/1538-4357/aaac2f}{ApJ},
  \hypersetup{urlcolor=blue}\href{https://ui.adsabs.harvard.edu/abs/2018ApJ...855....2Q}{855,
  2}

\bibitem[{{Schlafly} \& {Finkbeiner}(2011)}]{Schlafly_Finkbeiner_2011}
{Schlafly}, E.~F., \& {Finkbeiner}, D.~P. 2011,
  \hypersetup{urlcolor=magenta}\href{https://dx.doi.org/10.1088/0004-637X/737/2/103}{ApJ},
  \hypersetup{urlcolor=blue}\href{https://ui.adsabs.harvard.edu/abs/2011ApJ...737..103S}{737,
  103}

\bibitem[{{Smirnov}(1948)}]{KS_test}
{Smirnov}, N. 1948,
  \hypersetup{urlcolor=magenta}\href{https://dx.doi.org/10.1214/aoms/1177730256}{The
  Annals of Mathematical Statistics}, 19, 279

\bibitem[{Spearman(1904)}]{Spearman_1904}
Spearman, C. 1904, The American Journal of Psychology, 15, 72

\bibitem[{{Villar} {et~al.}(2018){Villar}, {Nicholl}, \&
  {Berger}}]{Villar_2018}
{Villar}, V.~A., {Nicholl}, M., \& {Berger}, E. 2018,
  \hypersetup{urlcolor=magenta}\href{https://dx.doi.org/10.3847/1538-4357/aaee6a}{\apj},
  \hypersetup{urlcolor=blue}\href{https://ui.adsabs.harvard.edu/abs/2018ApJ...869..166V}{869,
  166}

\bibitem[{{Villar} {et~al.}(2020){Villar}, {Hosseinzadeh}, {Berger},
  {Ntampaka}, {Jones}, {Challis}, {Chornock}, {Drout}, {Foley}, {Kirshner},
  {Lunnan}, {Margutti}, {Milisavljevic}, {Sanders}, {Pan}, {Rest}, {Scolnic},
  {Magnier}, {Metcalfe}, {Wainscoat}, \& {Waters}}]{Villar_2020}
{Villar}, V.~A., {Hosseinzadeh}, G., {Berger}, E., {et~al.} 2020,
  \hypersetup{urlcolor=magenta}\href{https://dx.doi.org/10.3847/1538-4357/abc6fd}{\apj},
  \hypersetup{urlcolor=blue}\href{https://ui.adsabs.harvard.edu/abs/2020ApJ...905...94V}{905,
  94}

\bibitem[{{Virtanen} {et~al.}(2020){Virtanen}, {Gommers}, {Oliphant},
  {Haberland}, {Reddy}, {Cournapeau}, {Burovski}, {Peterson}, {Weckesser},
  {Bright}, {van der Walt}, {Brett}, {Wilson}, {Millman}, {Mayorov}, {Nelson},
  {Jones}, {Kern}, {Larson}, {Carey}, {Polat}, {Feng}, {Moore}, {Vand erPlas},
  {Laxalde}, {Perktold}, {Cimrman}, {Henriksen}, {Quintero}, {Harris},
  {Archibald}, {Ribeiro}, {Pedregosa}, {van Mulbregt}, \& {SciPy 1. 0
  Contributors}}]{Virtanen_2020}
{Virtanen}, P., {Gommers}, R., {Oliphant}, T.~E., {et~al.} 2020,
  \hypersetup{urlcolor=magenta}\href{https://dx.doi.org/10.1038/s41592-019-0686-2}{NatMe},
  \hypersetup{urlcolor=blue}\href{https://ui.adsabs.harvard.edu/abs/2020NatMe..17..261V}{17,
  261}

\bibitem[{{Woosley}(2010)}]{Woosley_2010}
{Woosley}, S.~E. 2010,
  \hypersetup{urlcolor=magenta}\href{https://dx.doi.org/10.1088/2041-8205/719/2/L204}{ApJL},
  \hypersetup{urlcolor=blue}\href{https://ui.adsabs.harvard.edu/abs/2010ApJ...719L.204W}{719,
  L204}

\end{thebibliography}
